\documentclass{mn2e}
\usepackage{psfig}
\usepackage[authoryear]{natbib}

\newcommand\ein{{\it Einstein\ } }
\newcommand\ros{{\it ROSAT}}
\newcommand\asca{{\it ASCA}}
\newcommand\sax{{\it BEPPOSAX}}

\newcommand\xmm{{\it XMM-Newton}}
\newcommand\chandra{{\it Chandra}}

\newcommand\ecs{erg cm$^{-2}$ s$^{-1}$}
\newcommand\nh{N_{H}}

\title{The XMM-SSC 
survey of hard-spectrum \xmm\ sources 1: optically bright sources}
\author[Page, et al.]{
M.J. Page\(^{1}\),
I. Lehmann\(^{2}\),
Th. Boller\(^{2}\), 
M.G. Watson\(^{3}\), 
T. Dwelly\(^{1,4}\),
S. Hess\(^{2}\),
\and
I. Matute\(^{2}\),
N.S. Loaring\(^{1,5}\),
S. Rosen\(^{1}\),
H. Ziaeepour\(^{1}\),
A. Schwope\(^{6}\),
\and
G. Lamer\(^{6}\),
F.J. Carrera\(^{7}\),
J. Tedds\(^{3}\),
R. Della Ceca\(^{8}\),
P. Severgnini\(^{8}\),
\and
R.G. McMahon\(^{9}\),
and
W. Yuan\(^{9,10}\)
\\ \\
\(^{1}\)Mullard Space Science Laboratory, University College London,
Holmbury St Mary, Dorking, Surrey RH5 6NT, UK.\\
\(^{2}\)Max-Planck-Institut f\"ur extraterrestrische Physik, 
Giessenbachstrasse, Postfach 1312, 85741 Garching, Germany.\\
\(^{3}\)Department of Physics and Astronomy, University of Leicester, 
LE1 7RH, UK.\\
\(^{4}\)School of Physics and Astronomy, University of Southampton, 
Southampton, SO17 1BJ, UK\\
\(^{5}\)South African Astronomical Observatory, Observatory, 
7935 Cape Town, South Africa\\
\(^{6}\)Astrophysikalisches Institut Potsdam (AIP), An der Sternwarte 16, 
14482 Potsdam, Germany.\\
\(^{7}\)Instituto de F\'\i sica de Cantabria 
(CSIC--Universidad de Cantabria), 39005
Santander, Spain.\\
\(^{8}\)INAF-Osservatorio Astronomico di Brera, via Brera 28, 
I-20121 Milano, Italy.\\
\(^{9}\)University of Cambridge, Institute of Astronomy, Madingley Road, 
Cambridge CB3 0HA, UK\\
\(^{10}\)National Astronomical Observatory of China/Yunnan Observatory, 
Kunming, 650011, PO Box 110 Yunnan, China\\
}

\date{}

\begin{document}
\maketitle

\begin{abstract}
We present optical and X-ray data for a sample of  
serendipitous \xmm\ sources
that are selected to have 0.5-2 keV vs 2-4.5 keV X-ray 
hardness ratios which are harder than the X-ray background. 
The sources have 2-4.5 keV X-ray flux $\geq 10^{-14}$ \ecs, and in this paper
we examine a subsample of 42 optically bright (r $<$ 21) sources; this
subsample is 100 per cent spectroscopically identified.
All but one of the optical counterparts are extragalactic, and we argue 
that the single
exception, a Galactic M star, is probably a coincidental association rather
than the correct identification of the X-ray source.
The X-ray spectra of all the sources are consistent with heavily absorbed
power laws ($21.8 < \log \nh < 23.4$), 
and all of them, including the 2
sources with 2-10~keV intrinsic luminosities of
$<10^{42}$~erg~s$^{-1}$, appear to be absorbed AGN. 
The majority of the sources show only
narrow emission lines in their optical spectra, implying that they are type-2
AGN. Three sources have 2-10~keV luminosities of
$>10^{44}$~erg~s$^{-1}$, and two of these sources have optical spectra which
are dominated by narrow emission lines, i.e. are type-2 QSOs.
Only a small fraction of the sources (7/42) show broad optical emission
lines, and all of these have $\nh < 10^{23}$\,cm$^{-2}$. 
%This implies that cold column densities of $> 10^{23}$~cm$^{-2}$ are rare in
%broad line objects with r $<$ 21 (at most a few percent of the population). 
This implies
that ratios of X-ray absorption to optical/UV extinction equivalent to $> 100
\times$ the Galactic gas-to-dust ratio are rare in AGN absorbers 
(at most a few percent of the population), and may be restricted to 
broad absorption-line QSOs.
Seven objects appear to have an additional soft X-ray component in
addition to the heavily absorbed power law; all seven are narrow emission line
objects with $z<0.3$ and 2-10~keV intrinsic luminosities 
$<10^{43}$~erg~s$^{-1}$. 
We consider the implications of our results in the light of the AGN unified
scheme. We find that the soft components in narrow-line objects are consistent
with the unified scheme provided that $> 4$ per cent of broad-line AGN have
ionised absorbers that attenuate their soft X-ray flux by $>50$ per cent.
In at least one of the X-ray absorbed, broad-line AGN in our sample the 
X-ray spectrum requires an ionised absorber, consistent with this picture.

%implications
%

\end{abstract}

\section{Introduction}

Deep surveys with \chandra\ and \xmm\ have now resolved the majority of 
the extragalactic X-ray background
(XRB) at $<5$ keV into
point sources \citep{alexander03,rosati02,worsley04,bauer04}. 
At the bright flux limits that were probed with \ein\ and \ros\ surveys,
 the soft X-ray sky is dominated by broad-line AGN
with soft X-ray spectra \citep{maccacaro88,mittaz99}.
At faint fluxes, the bulk of the sources have
hard X-ray spectra \citep{giacconi01,brandt01}, 
as required to produce the overall
XRB spectral shape. XRB synthesis models predict that
these faint, hard sources are intrinsically absorbed 
AGN \citep[e.g.][]{setti89,comastri95,gilli01}. 
These absorbed
sources are likely to be the dominant AGN population: they are responsible 
for the
majority of the XRB energy density and probably outnumber
unabsorbed AGN by a factor of 4 or more \citep{fabianiwasawa99}, as is found 
in optical and X-ray studies of local AGN \citep{maiolino95,risaliti99}.
Optical/infrared imaging and spectroscopy 
of the faint, hard-spectrum sources, while consistent with 
the absorbed AGN hypothesis, reveals a heterogeneous
population. Many of these sources are narrow-line AGN (Seyfert 2s and
QSO 2s), others
show no emission lines
whatsoever, while some are optically
unobscured, broad line 
AGN \citep[e.g.][]{mainieri02,barger03,szokoly04,page06b}. 
This variety suggests that there is a large range of gas-to-dust
ratios within the absorbed population.

There are a variety of important issues regarding the absorbed population 
which remain to be addressed. First, the relationship between the 
optical spectroscopic characteristics and the level of X-ray absorption is not
fully understood 
\citep[e.g.][]{maiolino01,page01,boller03,barcons03a,caccianiga04,carrera04}.
Nor is it established whether the 
fraction of absorbed AGN 
depends on redshift and/or on luminosity: several recent studies suggest that 
the fraction of absorbed AGN decreases with luminosity and/or redshift 
\citep[e.g.][]{cowie03,barger03,steffen03,ueda03,lafranca05}, 
while other studies 
suggest that the 
absorbed fraction is independent of luminosity 
\citep[e.g.][]{treister04,dwelly05,dwelly06,tozzi06}. 
Indeed, this issue has 
been debated since at least the early 1980s 
\citep[e.g.][]{lawrence82,mushotzky82,maccacaro84}. 
Another particularly important question is whether the absorbed AGN 
population is wholly related to the unabsorbed population by 
AGN geometric unification
schemes \citep{antonucci93}. Broad line AGN with significant 
X-ray absorption but little or no optical/UV obscuration 
\citep[e.g.][]{barr77,elvis94,akiyama00,page01}
have proven
particularly challenging to fit within such schemes.
Alternatively, parts of the absorbed 
population may be
physically distinct from the unabsorbed population, 
as predicted by some evolutionary 
models \citep{sanders88,fabian99,franceschini02,elvis02,page04}.

In this paper we present a sample of serendipitous \xmm\ sources selected to
have hard X-ray spectra in the 0.5-4.5 keV energy range. This sample is
well suited for the study of the X-ray and optical characteristics of the 
X-ray/optically bright part of the X-ray absorbed AGN population.
We describe the X-ray and optical selection of the sample in detail 
in Section \ref{sec:selection}, and in Sections \ref{sec:opticalobs} and 
\ref{sec:xrayspectra} we describe the optical spectroscopic observations and
the construction of the X-ray spectra. Our results are presented in Section
\ref{sec:results} and their implications for the absorbed AGN population are
discussed in Section \ref{sec:discussion}. Throughout this paper we assume
$H_{0}=70$~km~s$^{-1}$~Mpc$^{-1}$, $\Omega_{m}=0.3$, and 
$\Omega_{\Lambda}=0.7$. We define a power law spectrum such that 
$dN/dE = AE^{-\Gamma}$
where $N$ is the number of photons, $E$ is photon energy, 
$\Gamma$ is the photon index and 
$A$ is the normalisation.  

\section{Sample selection}
\label{sec:selection}

Thanks to \xmm's exceptional throughput and large field of view (30$\arcmin$
diameter), a serendipitous \xmm\ catalogue is the obvious resource with which
to find bright examples of the absorbed AGN population. Our serendipitous
catalogue was constructed by concatenating the source lists from
100 pipeline processed \xmm\ fields. This catalogue preceded the public release
of the 1XMM catalogue, which is about 6 times larger, \citep{watson03} but
contains similar information about each source.
The targets of the \xmm\ observations were excluded from the catalogue.
Hard-spectrum
sources are heavily outnumbered at bright fluxes by soft-spectrum, unabsorbed
AGN, even in the 2-10 keV band. Therefore we have selected our sample from 
the catalogue using the 2-4.5 keV vs 0.5-2 keV hardness ratio. We refer to
this ratio as ``$HR2$'' to follow the convention used in the 1XMM catalogue.
$HR2$ is defined as
\begin{equation}
HR2 = \frac{CR_{2-4.5}-CR_{0.5-2}}{CR_{2-4.5}+CR_{0.5-2}}
\end{equation}
where $CR_{2-4.5}$ is the vignetting-corrected countrate in the 2-4.5 keV range
and $CR_{0.5-2}$ is the vignetting-corrected countrate in the 0.5-2 keV
range. $HR2$ is calculated independently for each of the 3 EPIC cameras.
The specific criterion used to select the sample is:
\begin{equation}
HR2 - \sigma_{HR2} > -0.3
\label{eq:selection}
\end{equation}
where $\sigma_{HR2}$ is the 68\% uncertainty on $HR2$. 
%comparison with 13hr hardness ratios - these are the important sources
The $HR2$ threshold of $-0.3$ corresponds approximately to the slope of the XRB
in the same energy range, so that only objects with spectra harder than the XRB
will be picked up with this criterion. This range of $HR2$ also corresponds to
the most heavily absorbed AGN typically found in \xmm\ surveys 
\citep{caccianiga04,dellaceca04}, the
numbers and redshifts of which place significant constraints on XRB synthesis
models \citep{dwelly05}.
For inclusion in the
sample, we require a source to have a detection likelihood 
(DET\_ML in 1XMM) $> 7.5$\footnote{A coding error in the task {\sc
emldetect} led to overestimated values for DET\_ML prior to SAS version
6.0. The values of DET\_ML given in this paper have been corrected 
for this
problem.} in the 2-4.5 keV band, 2-4.5 keV flux 
$>10^{-14}$~erg~cm$^{-2}$~s$^{-1}$ 
Galactic $\nh < 5\times 10^{20} {\rm cm}^{-2}$, $|b| >
30$ degrees, and 
satisfy equation \ref{eq:selection} in at least one  
of the three EPIC cameras. The DET\_ML threshold of 7.5 corresponds to a
probability of being a background fluctuation of $5\times 10^{-4}$, and a
spurious source fraction of $<1\%$ \citep{watson03}.
%according to theory, we'd expect about 1% spurious sources at the threshold 
%of ML=7.5, but according to Figure 5.3, of watson03 the fraction is < 1%
%in the multi-ML pass.
From a total of 9340 sources with DET\_ML~$>~7.5$ found in the 100 
{\em XMM-Newton} fields used here, we obtained a sample 
of 136 sources which meet the selection criteria for our hard spectrum sample.

\begin{figure}
\begin{center}
\leavevmode
\psfig{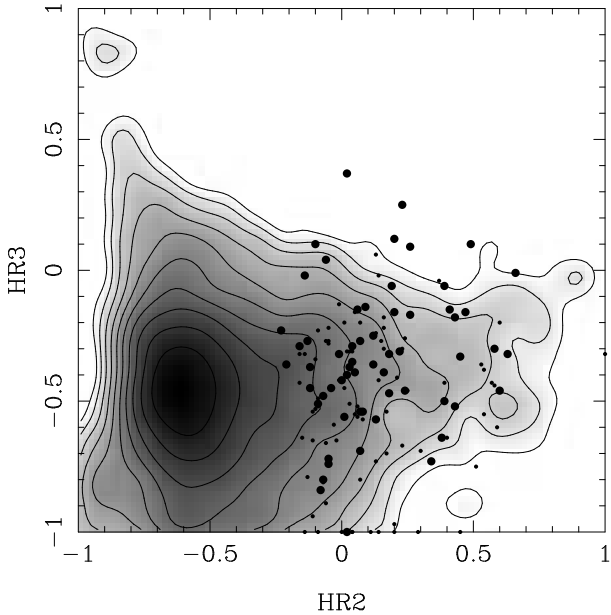}
\caption{$HR2$ and $HR3$ hardness ratio distribution of the optically bright
sources (large
dots) and optically-faint / blank-field sources (small dots) from the hard
spectrum sample. The underlying greyscale and contours show the pn-camera 
hardness ratio
distribution of the entire high galactic latitude ($\mid b\mid >20^{\circ}$)
 1XMM catalogue, and have a logarithmic scaling \citep{watson03}.}
\label{fig:HR_1_2_grey}
\end{center}
\end{figure}

The sources were categorised according to optical magnitude to 
facilitate our optical identification programme. Those sources with 
counterparts on SuperCOSMOS digitisations of the Palomar or 
UK Schmidt Telescope 
sky survey plates within 6$\arcsec$ of the X-ray position 
will be referred to as the optically bright sample; this amounts to 
73 sources, and provides
the target list for our spectroscopic 
identification programme on 4m telescopes. Those sources without sky 
survey counterparts\footnote{We take the the sky survey plate
limit to be $r \sim 21$, and consider sources with $r > 21$ as part of the
optically-faint sample even if they are detected on the sky survey plates.}
but
which have counterparts in deeper imaging form the target list for
our spectroscopic identification programme on 8-10m telescopes and are
considered the optically-faint sample. Our identification programme for
the optically faint sample is ongoing. The final category of sources 
(``blank-field sources'') are those for which no optical counterpart is 
found to $r \sim 23$. 
Discussion of the optically faint and blank-field samples is deferred to a
future paper pending the results of the 8-10m spectroscopic campaign.

The hardness ratios, $HR2$ and $HR3$, for the full sample of hard spectrum
sources are shown in Fig. \ref{fig:HR_1_2_grey}, where $HR3$ is the 4.5-7.5 keV
vs 2-4.5 keV hardness ratio, defined as
\begin{equation}
HR3 = \frac{CR_{4.5-7.5}-CR_{2-4.5}}{CR_{4.5-7.5}+CR_{2-4.5}}
\end{equation}
The optically bright sources are shown as large dots, while the optically 
faint and blank-field sources are shown as small dots. 
The optically faint and
optically bright parts of the sample have similar hardness ratio 
distributions: the two distributions are indistinguishable at the 90 per cent
confidence level according to the  2-dimensional Kolmogorov-Smirnov test 
\citep{fasano87}.
The underlying greyscale and contours show the hardness ratio
distribution for the $\mid b\mid >20^{\circ}$ 1XMM catalogue \citep{watson03}. 
The peak in the underlying
distribution at ($-0.6$,$-0.5$) corresponds to the typical spectra of
unabsorbed, broad-line AGN. While such sources dominate the 1XMM catalogue,
(and all medium-depth, flux-limited samples of serendipitous {\em XMM-Newton} 
sources) 
the distribution shows a significant tail stretching to larger values of 
$HR2$. 
The hard spectrum sample described in this paper follows the underlying
distribution of 1XMM sources well for $HR2>$$-0.2$, and so is ideal for the study
of this hard-spectrum tail of the population.

\begin{figure}
\begin{center}
\leavevmode
\psfig{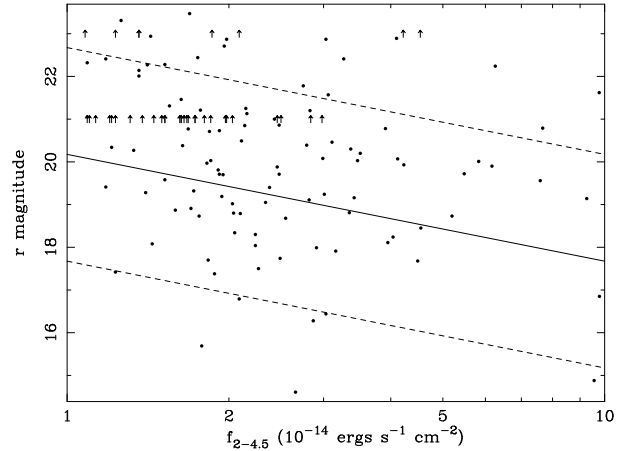}
\caption{2-4.5 keV X-ray flux vs optical $r$-band magnitude 
for the hard-spectrum sources. The
solid line corresponds to 
the ratio $f_{X}/f_{R} = 1$ and the dashed lines indicate
the $\pm 1$~dex interval, where $f_{X}/f_{R}$ is 
the ratio of X-ray to optical flux as defined in \citet{mchardy03}. The 2 - 4.5
keV fluxes ($f_{2-4.5}$) were determined from the 2-4.5~keV count rates, 
assuming a
photon index $\Gamma = 1.7$.}
\label{fig:X_R}
\end{center}
\end{figure}

The optical magnitudes and X-ray fluxes of the hard
spectrum sources are shown in Fig. \ref{fig:X_R}, and in this figure we also
show the interval $\mid \log_{10} (f_{X}/f_{R}) \mid < 1$, where $f_{X}$
and $f_{R}$ are X-ray flux and optical flux respectively, as defined by
\citet{mchardy03}\footnote{The relationship $f_{X}=f_{R}$ as defined by
\citep{mchardy03} translates to $r = 20.175 - \log f_{2-4.5}$ where $f_{2-4.5}$
is the 2 - 4.5\,keV flux in units of $10^{-14}$~ergs~cm$^{-2}$~s$^{-1}$,
assuming an X-ray photon index $\Gamma=1.7$.}. The majority of 
hard-spectrum sources
lie within this interval, as do most of the X-ray selected population at 
this flux level \citep{stocke91}. 

\section{Optical spectroscopic observations}
\label{sec:opticalobs}

\begin{figure*}
\vspace{-1.6cm}
\begin{center}
\leavevmode
\psfig{figure=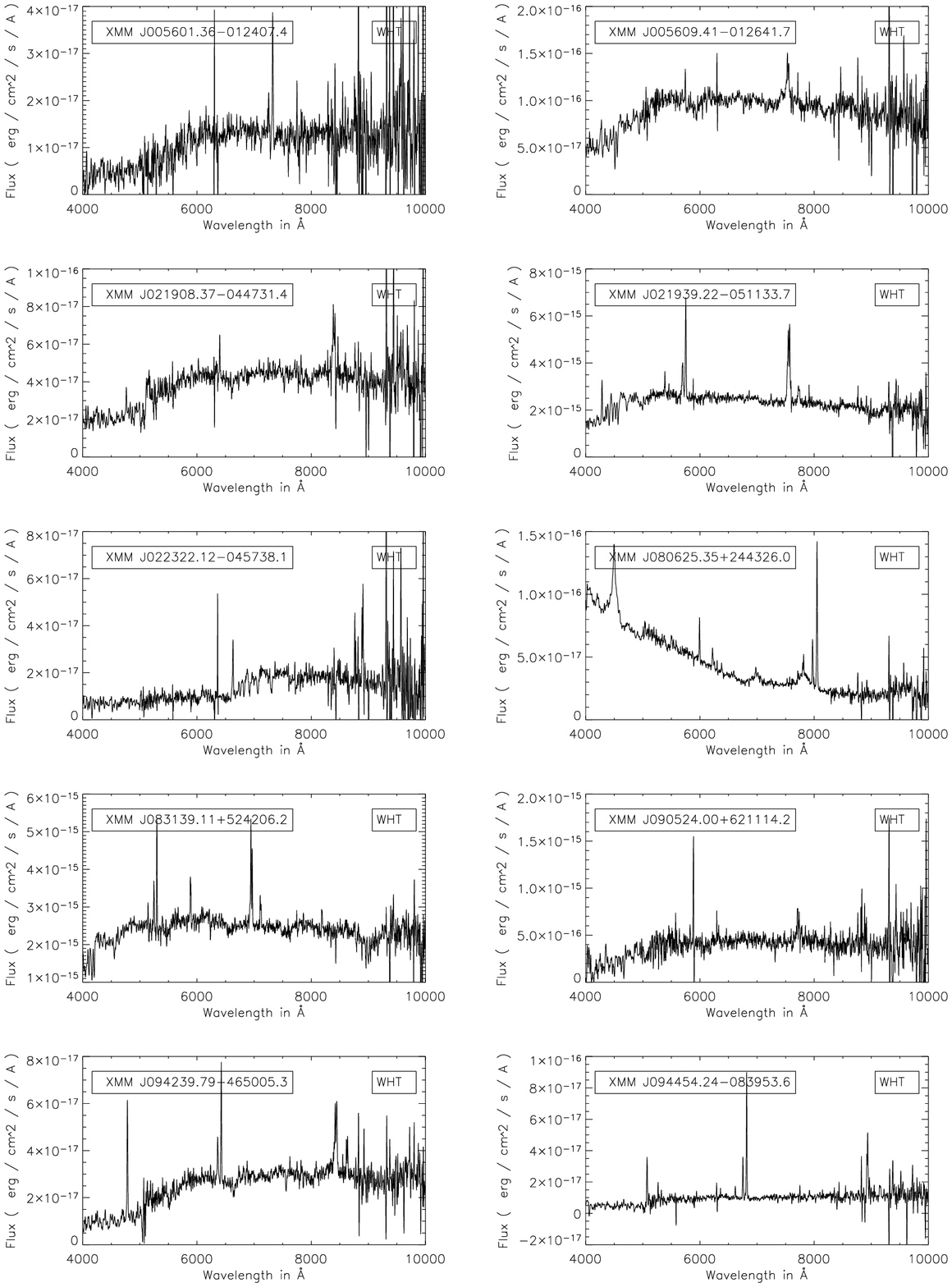,width=180truemm}
\vspace{-3cm}
\caption{Optical spectra of hard-spectrum serendipitous sources. }
\label{fig:optspectra}
\end{center}
\end{figure*}
\begin{figure*}
\vspace{-1.6cm}
\begin{center}
\leavevmode
\psfig{figure=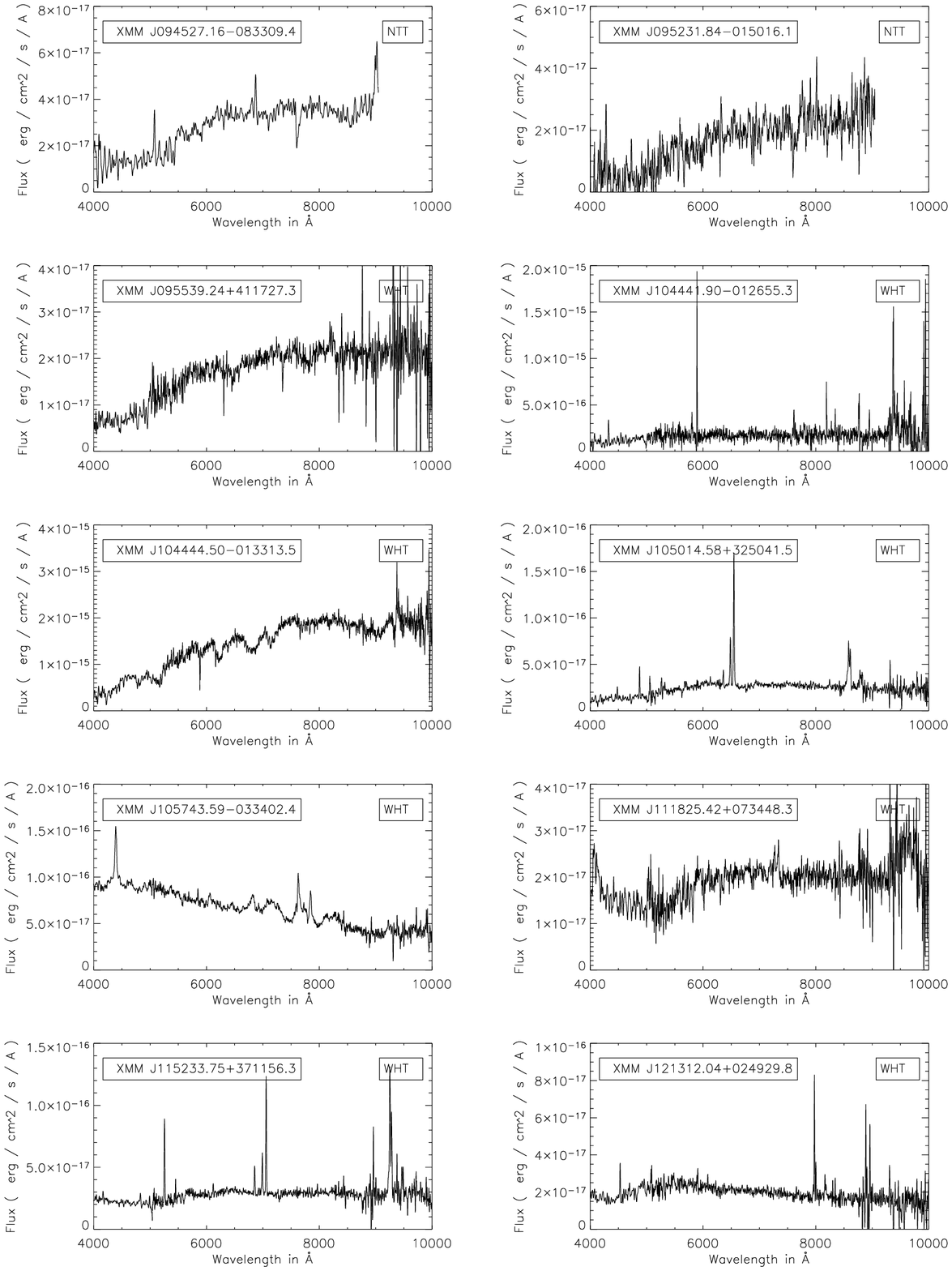,width=180truemm}
\vspace{-3cm}
\end{center}
{\bf Figure \ref{fig:optspectra}} {\it continued}
\end{figure*}
\begin{figure*}
\vspace{-1.6cm}
\begin{center}
\leavevmode
\psfig{figure=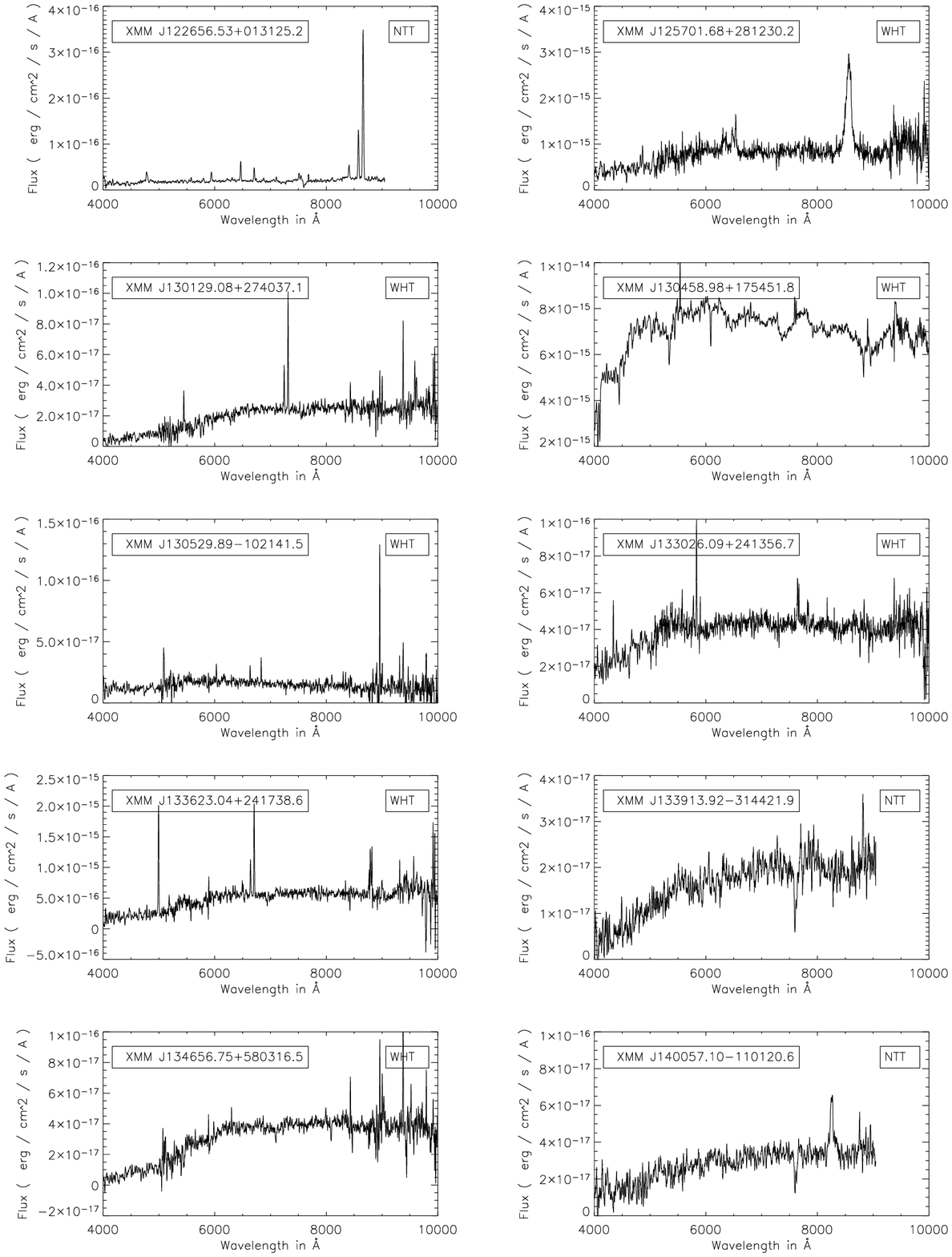,width=180truemm}
\vspace{-3cm}
\end{center}
{\bf Figure \ref{fig:optspectra}} {\it continued}
\end{figure*}
\begin{figure*}
\vspace{-1.6cm}
\begin{center}
\leavevmode
\psfig{figure=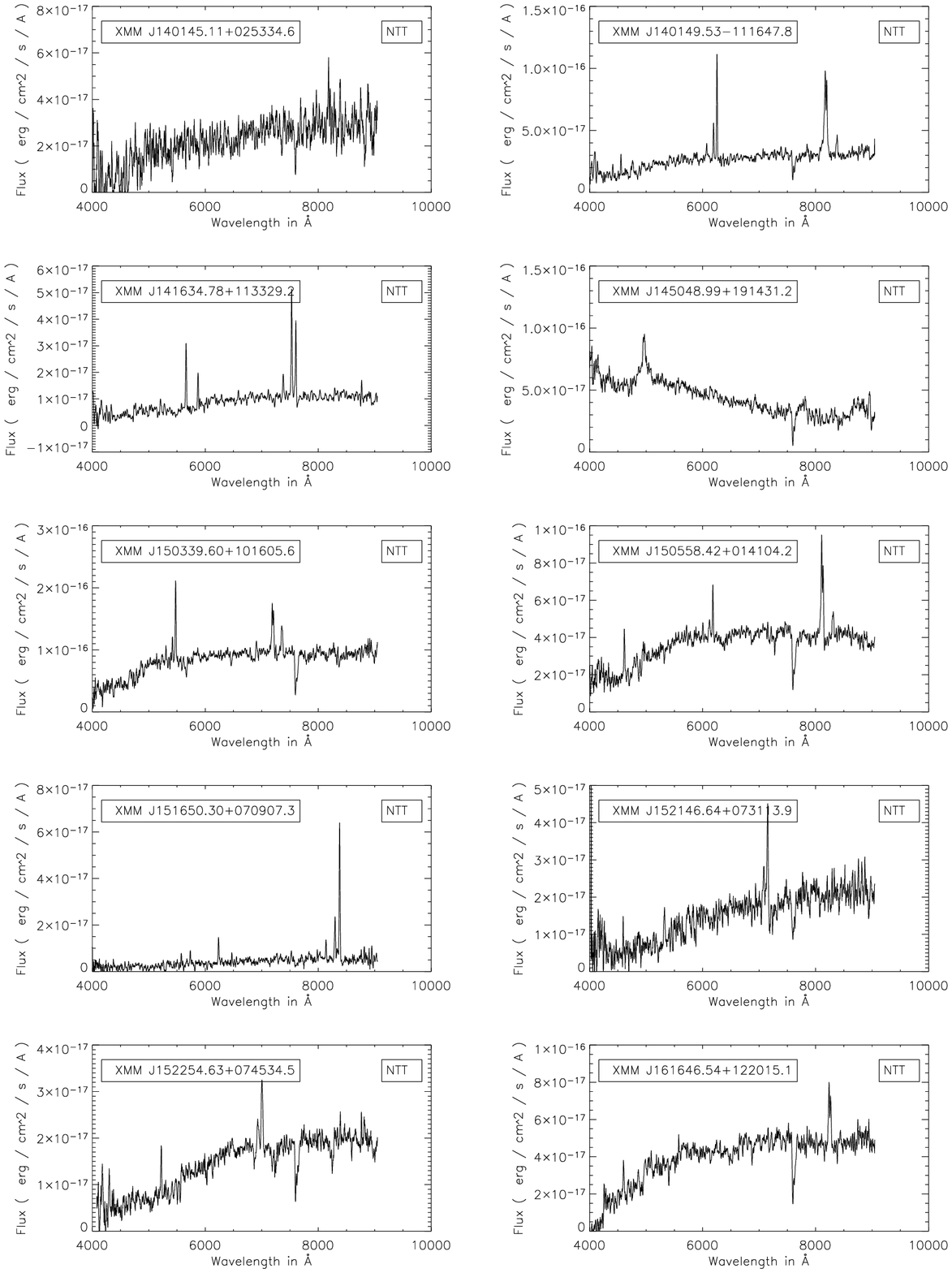,width=180truemm}
\vspace{-3cm}
\end{center}
{\bf Figure \ref{fig:optspectra}} {\it continued}
\end{figure*}
\begin{figure*}
\vspace{-9.6cm}
\begin{center}
\leavevmode
\psfig{figure=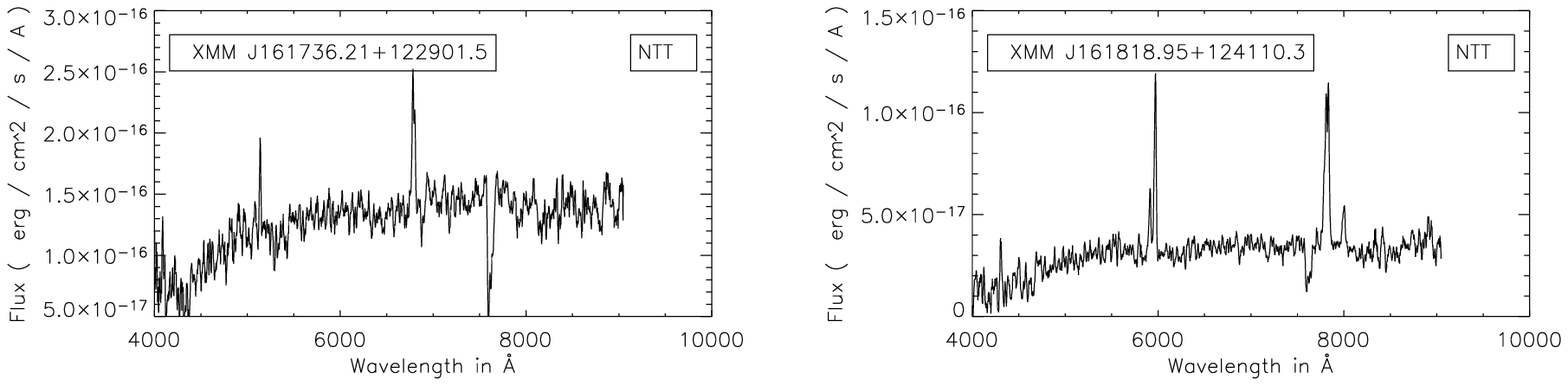,width=180truemm}
\vspace{-11.4cm}
\end{center}
{\bf Figure \ref{fig:optspectra}} {\it continued}
\end{figure*}

Optical spectra were obtained in two observing runs. The first run took place
on the nights of 27th and 28th January 2003 at the William Herschel Telescope on La Palma. Observing
conditions were variable, but most of the data were taken in conditions of good
seeing ($\sim 1\arcsec$) and sky transparency. Observations
were made with the ISIS dual arm spectrograph using a 1.5$''$ slit oriented at
the parallactic angle, and 
employing the 5400\,\AA\
dichroic. On the blue arm we used the R300B grating, covering the wavelength
range from 3000 -- 5700\,\AA\ at 4.9\,\AA\ 
%% measured from arc lines taken through the slit
resolution (measured from arc lines taken through the slit) and on the red arm we used 
the R158R grating, covering the wavelength range 5000 -- 10000\,\AA\ at 10.1\,\AA\ 
resolution. Spectra of CuNe and CuAr arclamps were used to calibrate the
wavelength scale. Flux calibration, and approximate correction for the telluric
absorption bands at $\sim 6900$ and $\sim 7600$\,\AA\ were achieved through 
observations of 
G191-B2B and BD+33 2642. Data were reduced using standard {\small IRAF} 
routines.
The reduced, flux calibrated spectra from the red and blue arms were then 
merged to form a single spectrum for each target.

The second observing run took place at the New Technology Telescope (NTT) at 
La Silla on the nights of the 2nd and 3rd May 2003 in conditions of variable seeing and sky transparency. 
Observations were made using the EMMI spectrograph using
a $1''$ slit at the parallactic angle and grism $\#$3, providing 8\,\AA\ resolution and a wavelength
range of 3800--9000\,\AA. Spectra of He and Ar arclamps were used to calibrate
the wavelength scale. Observations of the spectrophotometric standards 
LTT4364 and LTT2415 were used for relative flux calibration.

\begin{figure}
\begin{center}
\leavevmode
\psfig{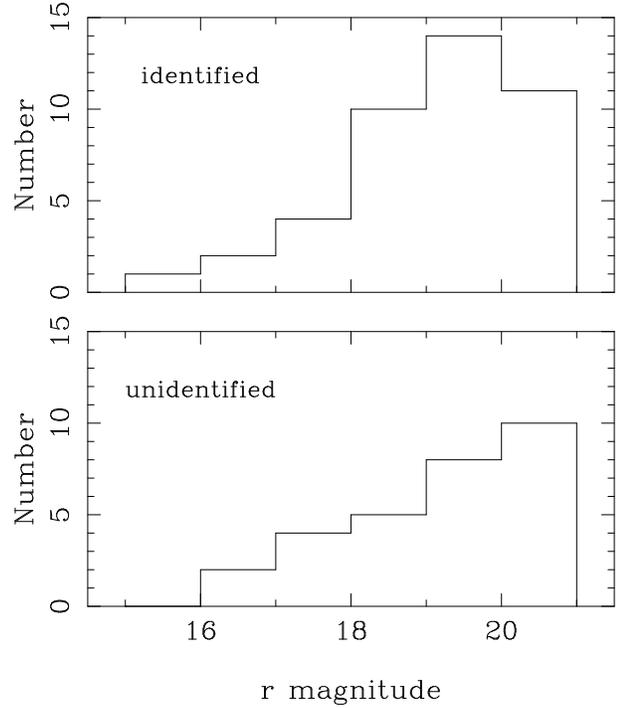}
\caption{Magnitude distributions for sources with r$<$21 which were (upper
panel) and were not (lower panel) observed during our optical spectroscopic
observations.}
\label{fig:brightness_histo}
\end{center}
\end{figure}

In total, spectra of the optical counterparts to 42 hard-spectrum sources were 
observed. The spectroscopic completeness is 100\% for the observed sample, 
i.e. an adequate spectrum yielding a reliable redshift was obtained for every 
optical counterpart that was observed. Apart from the selection of counterparts
from sky survey plates, we did not bias the spectroscopic observations to
bright sources. To demonstrate this, Fig. \ref{fig:brightness_histo} 
compares the
optical magnitude distributions for the optically bright sources which were and
were not observed spectroscopically; the two distributions are
indistinguishable according to the Kolmogorov-Smirnov test, which gives the 
probability that they were both drawn from the same distribution as 
46 per cent. Hence our spectroscopically identified sources are a 
representative and statistically-complete sample of the optically bright
(r$<21$), hard spectrum population. 

\section{Construction of X-ray spectra}
\label{sec:xrayspectra}

%All the target sources are \xmm\ serendipitous sources, and so all have some 
%X-ray spectral information available. Therefore 
We extracted X-ray 
spectra for every source using the available \xmm\ EPIC data. For sources which
are present in more than one \xmm\ observation in the public archive, we have
used all the available data to obtain the highest possible signal to noise 
ratio. Table \ref{tab:observations} gives the \xmm\ observations used to
construct the X-ray spectrum of each source. 
The EPIC data were reduced using the Science Analysis System (SAS)
verson 5.4. The \xmm\ point spread function changes with off-axis
angle. Therefore source spectra were extracted from elliptical source 
regions with
sizes and orientations depending on the fluxes and off-axis angles of the
sources. Typically, these regions have major axes of $\sim 18''$ and 
minor axes of $\sim 10''$. Background spectra were extracted from annular 
regions, of 2.5$'$ outer radius, centred on the source, and with all 
significant X-ray sources excised.
Response matrices and the appropriate effective area files were computed using
the SAS tasks {\small RMFGEN} and {\small ARFGEN} respectively. For each source
the EPIC spectra were combined to form a single spectrum using the method of
\citet{page03}. Source spectra were grouped using the HEASOFT tool 
{\small GRPPHA} to
ensure the minimum number of counts per bin given in Table \ref{tab:nhfits}
(usually 20) before fitting.

\begin{table*}
\begin{center}
\caption{\xmm\ observations used to construct the X-ray spectra.}
\label{tab:observations}
\begin{tabular}{l@{\hspace{7mm}}lll}
Source&\multicolumn{3}{c}{--- \xmm\ observation IDs ---}\\
&&&\\
XMM~J$005601.36$$-$$012407.4$ & 0012440101 &            &	     \\
XMM~J$005609.41$$-$$012641.7$ & 0012440101 &            &	     \\
XMM~J$021908.37$$-$$044731.4$ & 0112370401 & 0112371501 & 0112370301 \\
XMM~J$021939.22$$-$$051133.7$ & 0112370301 &            &	     \\
XMM~J$022322.12$$-$$045738.1$ & 0109520501 &            &	     \\
XMM~J$080625.35$$+$$244326.0$ & 0094530401 &            &	     \\
XMM~J$083139.11$$+$$524206.2$ & 0092800201 & 0092800101 &	     \\
XMM~J$090524.00$$+$$621114.2$ & 0110661701 & 0110660201 &	     \\
XMM~J$094239.79$$+$$465005.3$ & 0106460101 &            &	     \\
XMM~J$094454.24$$-$$083953.6$ & 0017540101 &            &	     \\
XMM~J$094527.16$$-$$083309.4$ & 0017540101 &            &	     \\
XMM~J$095231.84$$-$$015016.1$ & 0065790101 &            &	     \\
XMM~J$095539.24$$+$$411727.3$ & 0111290201 &            &	     \\
XMM~J$104441.90$$-$$012655.3$ & 0125300101 &            &	     \\
XMM~J$104444.50$$-$$013313.5$ & 0125300101 &            &	     \\
XMM~J$105014.58$$+$$325041.5$ & 0055990201 &            &	     \\
XMM~J$105743.59$$-$$033402.4$ & 0094800101 &            &	     \\
XMM~J$111825.42$$+$$073448.3$ & 0082340101 &            &	     \\
XMM~J$115233.75$$+$$371156.3$ & 0112551401 &            &	     \\
XMM~J$121312.04$$+$$024929.8$ & 0081340801 &            &	     \\
XMM~J$122656.53$$+$$013125.2$ & 0110990201 &            &	     \\
XMM~J$125701.68$$+$$281230.2$ & 0124710701 & 0124712101 &	     \\
XMM~J$130129.08$$+$$274037.1$ & 0124710801 &            &	     \\
XMM~J$130458.98$$+$$175451.8$ & 0017940101 &            &	     \\
XMM~J$130529.89$$-$$102141.5$ & 0032141201 &            &	     \\
XMM~J$133026.09$$+$$241356.7$ & 0100240101 & 0100240201 &	     \\
XMM~J$133623.04$$+$$241738.6$ & 0096010101 &            &	     \\
XMM~J$133913.92$$-$$314421.9$ & 0035940301 &            &	     \\
XMM~J$134656.75$$+$$580316.5$ & 0112250201 &            &	     \\
XMM~J$140057.10$$-$$110120.6$ & 0109910101 &            &	     \\
XMM~J$140145.11$$+$$025334.6$ & 0098010101 &            &	     \\
XMM~J$140149.53$$-$$111647.8$ & 0109910101 &            &	     \\
XMM~J$141634.78$$+$$113329.2$ & 0112250301 & 0112251301 &	     \\
XMM~J$145048.99$$+$$191431.2$ & 0056030101 &            &	     \\
XMM~J$150339.60$$+$$101605.6$ & 0112910101 & 0070740101 & 0070740301 \\
XMM~J$150558.42$$+$$014104.2$ & 0021540101 &            &	     \\
XMM~J$151650.30$$+$$070907.3$ & 0109920301 & 0109920101 &	     \\
XMM~J$152146.64$$+$$073113.9$ & 0109930101 &            &	     \\
XMM~J$152254.63$$+$$074534.5$ & 0109930101 &            &	     \\
XMM~J$161646.54$$+$$122015.1$ & 0103460901 & 0103460801 &	     \\
XMM~J$161736.21$$+$$122901.5$ & 0103461001 & 0103460901 &	     \\
XMM~J$161818.95$$+$$124110.3$ & 0103461001 &            &	     \\
\end{tabular}
\end{center}
\end{table*}

\section{Results}
\label{sec:results}

\begin{table*}
\begin{center}
\caption{The X-ray sources and optical counterparts. X-ray fluxes are for the
2-4.5 keV band, in units
of $10^{-14}$~erg~cm$^{-2}$~s$^{-1}$. The column labelled `r'
gives the r magnitude from our Isaac Newton Telescope Wide Field Camera
images, except for those sources marked with an `$*$', in which case the
magnitude is UK Schmidt R (for southern fields) or POSS-II R (for
northern fields). The `cam' column gives the EPIC camera from which the 
source was selected. The column labelled `$o-x$' gives the offset between the
X-ray and optical positions in arcseconds.}
\begin{tabular}{l@{\hspace{2mm}}c@{\hspace{2mm}}r@{\hspace{3mm}}c@{\hspace{3mm}}c@{\hspace{2mm}}cclcc@{\hspace{1mm}}c}
 Source    &Flux &$HR2$&cam&Galactic $\nh$&\multicolumn{2}{c}{optical position (J2000)}&o-x&\ \ r
 &type&z\\
                 &(2-4.5) &             &    &  ($10^{20}$~cm$^{-2}$)&        RA   &       dec   & ($\arcsec$)&        &        &       \\
&&&&&&&&&&\\
XMM~J$005601.36$$-$$012407.4$ & 2.4 & $-0.04\pm 0.10$ & PN  & 3.24 & 00 56 01.46 & $-$01 24 05.0 & 2.8&21.0$^{*}$ &  NELG   & 0.463\\
XMM~J$005609.41$$-$$012641.7$ & 2.4 & $ 0.58\pm 0.15$ & PN  & 3.29 & 00 56 09.56 & $-$01 26 41.7 & 2.2&17.7$^{*}$ &  NELG   & 0.148\\
XMM~J$021908.37$$-$$044731.4$ & 5.4 & $ 0.04\pm 0.10$ & M2  & 2.55 & 02 19 08.38 & $-$04 47 30.7 & 0.7&19.7       &  NELG   & 0.279\\
XMM~J$021939.22$$-$$051133.7$ & 1.8 & $ 0.26\pm 0.15$ & PN  & 2.59 & 02 19 39.08 & $-$05 11 33.4 & 2.1&17.7       &  NELG   & 0.151\\
XMM~J$022322.12$$-$$045738.1$ & 1.9 & $ 0.60\pm 0.10$ & PN  & 2.63 & 02 23 21.95 & $-$04 57 38.9 & 2.7&20.7$^{*}$ &  NELG   & 0.780\\
XMM~J$080625.35$$+$$244326.0$ & 5.8 & $ 0.13\pm 0.19$ & M1  & 3.94 & 08 06 25.33 & $+$24 43 24.1 & 1.9&20.0       &  BLAGN  & 0.608\\
XMM~J$083139.11$$+$$524206.2$ & 1.2 & $ 0.02\pm 0.14$ & PN  & 3.84 & 08 31 39.08 & $+$52 42 05.8 & 0.5&17.4       &  NELG   & 0.059\\
XMM~J$090524.00$$+$$621114.2$ & 4.0 & $ 0.43\pm 0.16$ & M2  & 4.47 & 09 05 23.75 & $+$62 11 09.3 & 5.2&18.2$^{*}$ &  NELG   & 0.175\\
XMM~J$094239.79$$+$$465005.3$ & 3.5 & $ 0.24\pm 0.11$ & M1  & 1.25 & 09 42 39.99 & $+$46 50 04.9 & 2.1&20.3       &  NELG   & 0.284\\
XMM~J$094454.24$$-$$083953.6$ & 3.3 & $ 0.18\pm 0.16$ & M1  & 3.60 & 09 44 54.02 & $-$08 39 52.8 & 3.4&20.3$^{*}$ &  NELG   & 0.362\\
XMM~J$094527.16$$-$$083309.4$ & 1.7 & $ 0.49\pm 0.21$ & PN  & 3.58 & 09 45 27.18 & $-$08 33 08.1 & 1.3&18.7$^{*}$ &  NELG   & 0.369\\
XMM~J$095231.84$$-$$015016.1$ & 5.2 & $-0.08\pm 0.10$ & PN  & 3.96 & 09 52 31.79 & $-$01 50 13.9 & 2.3&18.7$^{*}$ &  galaxy & 0.316\\
XMM~J$095539.24$$+$$411727.3$ & 2.8 & $ 0.06\pm 0.21$ & PN  & 1.16 & 09 55 39.38 & $+$41 17 29.6 & 2.8&19.1$^{*}$ &  NELG   & 0.249\\
XMM~J$104441.90$$-$$012655.3$ & 1.9 & $ 0.05\pm 0.07$ & PN  & 4.18 & 10 44 41.88 & $-$01 26 56.2 & 1.0&19.1$^{*}$ &  NELG   & 0.160\\
XMM~J$104444.50$$-$$013313.5$ & 2.0 & $-0.01\pm 0.09$ & PN  & 4.19 & 10 44 44.48 & $-$01 33 10.6 & 2.9&18.3$^{*}$ &  M star & 0.000\\
XMM~J$105014.58$$+$$325041.5$ & 1.8 & $-0.13\pm 0.14$ & PN  & 2.00 & 10 50 14.58 & $+$32 50 42.6 & 1.1&20.0       &  NELG   & 0.309\\
XMM~J$105743.59$$-$$033402.4$ & 2.0 & $ 0.18\pm 0.14$ & PN  & 3.59 & 10 57 43.68 & $-$03 34 02.0 & 1.4&19.0       &  BLAGN  & 0.567\\
XMM~J$111825.42$$+$$073448.3$ & 1.3 & $ 0.22\pm 0.11$ & PN  & 3.75 & 11 18 25.25 & $+$07 34 48.8 & 2.6&20.3       &  BLAGN  & 0.466\\
XMM~J$115233.75$$+$$371156.3$ & 9.2 & $ 0.07\pm 0.13$ & M1  & 1.88 & 11 52 33.80 & $+$37 11 56.5 & 0.6&19.1$^{*}$ &  NELG   & 0.411\\
XMM~J$121312.04$$+$$024929.8$ & 2.1 & $ 0.42\pm 0.34$ & PN  & 1.79 & 12 13 12.17 & $+$02 49 25.9 & 4.4&18.8$^{*}$ &  NELG   & 0.216\\
XMM~J$122656.53$$+$$013125.2$ & 7.5 & $-0.23\pm 0.06$ & PN  & 1.84 & 12 26 56.46 & $+$01 31 24.6 & 1.2&19.6$^{*}$ &  BLAGN  & 0.732\\
XMM~J$125701.68$$+$$281230.2$ & 3.3 & $ 0.12\pm 0.12$ & PN  & 0.91 & 12 57 01.72 & $+$28 12 30.2 & 0.5&18.8$^{*}$ &  BLAGN  & 0.305\\
XMM~J$130129.08$$+$$274037.1$ & 2.4 & $ 0.38\pm 0.15$ & M1  & 0.94 & 13 01 29.17 & $+$27 40 37.3 & 1.2&19.7$^{*}$ &  NELG   & 0.462\\
XMM~J$130458.98$$+$$175451.8$ & 2.8 & $ 0.66\pm 0.18$ & M1  & 2.15 & 13 04 59.01 & $+$17 54 54.4 & 2.6&16.3       &  NELG   & 0.034\\
XMM~J$130529.89$$-$$102141.5$ & 1.4 & $ 0.39\pm 0.14$ & PN  & 3.32 & 13 05 29.89 & $-$10 21 42.3 & 0.8&19.3$^{*}$ &  NELG   & 0.366\\
XMM~J$133026.09$$+$$241356.7$ & 2.5 & $ 0.04\pm 0.08$ & PN  & 1.16 & 13 30 26.05 & $+$24 13 56.1 & 0.8&18.7$^{*}$ &  NELG   & 0.166\\
XMM~J$133623.04$$+$$241738.6$ & 4.1 & $ 0.12\pm 0.12$ & M2  & 1.17 & 13 36 22.90 & $+$24 17 36.5 & 2.8&20.1       &  NELG   & 0.341\\
XMM~J$133913.92$$-$$314421.9$ & 2.3 & $ 0.01\pm 0.12$ & M1  & 3.85 & 13 39 13.94 & $-$31 44 22.4 & 0.6&19.4$^{*}$ &  galaxy & 0.378\\
XMM~J$134656.75$$+$$580316.5$ & 3.4 & $ 0.47\pm 0.10$ & PN  & 1.27 & 13 46 56.75 & $+$58 03 15.7 & 0.8&20.0       &  galaxy & 0.373\\
XMM~J$140057.10$$-$$110120.6$ & 2.9 & $-0.12\pm 0.11$ & M2  & 4.12 & 14 00 57.09 & $-$11 01 21.2 & 0.6&18.0$^{*}$ &  BLAGN  & 0.257\\
XMM~J$140145.11$$+$$025334.6$ & 2.2 & $-0.21\pm 0.09$ & PN  & 2.33 & 14 01 45.02 & $+$02 53 33.1 & 2.0&18.0$^{*}$ &  NELG   & 0.242\\
XMM~J$140149.53$$-$$111647.8$ & 1.9 & $ 0.20\pm 0.12$ & PN  & 4.54 & 14 01 49.52 & $-$11 16 48.4 & 0.6&19.8       &  NELG   & 0.248\\
XMM~J$141634.78$$+$$113329.2$ & 3.9 & $ 0.08\pm 0.11$ & PN  & 1.82 & 14 16 34.82 & $+$11 33 31.5 & 2.4&20.8       &  NELG   & 0.519\\
XMM~J$145048.99$$+$$191431.2$ & 3.9 & $ 0.02\pm 0.10$ & M2  & 2.48 & 14 50 48.98 & $+$19 14 30.9 & 0.3&18.1$^{*}$ &  BLAGN  & 0.774\\
XMM~J$150339.60$$+$$101605.6$ & 9.7 & $ 0.63\pm 0.10$ & M1  & 2.32 & 15 03 39.50 & $+$10 16 03.0 & 3.0&16.9$^{*}$ &  NELG   & 0.088\\
XMM~J$150558.42$$+$$014104.2$ & 3.4 & $ 0.16\pm 0.08$ & PN  & 4.26 & 15 05 58.38 & $+$01 41 04.3 & 0.6&19.2       &  NELG   & 0.237\\
XMM~J$151650.30$$+$$070907.3$ & 3.1 & $ 0.07\pm 0.14$ & M1  & 2.68 & 15 16 50.37 & $+$07 09 04.3 & 3.2&20.5$^{*}$ &  NELG   & 0.674\\
XMM~J$152146.64$$+$$073113.9$ & 4.2 & $-0.16\pm 0.06$ & PN  & 3.05 & 15 21 46.68 & $+$07 31 13.4 & 0.8&19.9       &  NELG   & 0.429\\
XMM~J$152254.63$$+$$074534.5$ & 1.1 & $-0.06\pm 0.20$ & PN  & 3.12 & 15 22 54.63 & $+$07 45 33.4 & 1.1&19.4$^{*}$ &  NELG   & 0.400\\
XMM~J$161646.54$$+$$122015.1$ & 2.3 & $-0.09\pm 0.14$ & PN  & 4.57 & 16 16 46.37 & $+$12 20 17.6 & 3.5&19.1$^{*}$ &  NELG   & 0.256\\
XMM~J$161736.21$$+$$122901.5$ & 3.3 & $ 0.09\pm 0.18$ & PN  & 4.59 & 16 17 36.25 & $+$12 29 03.3 & 1.9&15.7$^{*}$ &  NELG   & 0.030\\
XMM~J$161818.95$$+$$124110.3$ & 3.3 & $ 0.43\pm 0.28$ & M2  & 4.56 & 16 18 19.00 & $+$12 41 11.6 & 1.5&18.3$^{*}$ &  NELG   & 0.191\\
\end{tabular}
\label{tab:optids}
\end{center}
\end{table*}

\subsection{Optical spectra}

The optical spectra are shown in Fig. \ref{fig:optspectra}, and the
identifications, redshifts and emission line characteristics are given 
in Table \ref{tab:optids}. The distribution of offsets between the X-ray
sources and optical counterparts is similar to those found in other {\em
XMM-Newton} surveys of similar depth \citep[e.g. ][]{barcons02,fiore03}, with
90 per cent of optical counterparts lying 
within 3 arcseconds of the X-ray source.
Note that two of the sources,
XMM~J$122656.53$$+$$013125.2$ and XMM~J$134656.75$$+$$580316.65$ are also
members of
the \xmm\ Bright
Serendipitous Survey and are also reported in \citet{dellaceca04}, while one of
the sources, XMM~J$021939.22$$-$$051133.7$, is also a member of 
the \xmm\ Medium Sensitivity Survey and is also reported in \citet{barcons07}.
We have split the objects into four categories.
Objects which have any permitted emission lines with measured 
FWHM\footnote{The FWHM have not been corrected for the instrumental
contribution, which is up to 600~km~s$^{-1}$ at the blue end of EMMI.} 
$>$~1000~km~s$^{-1}$ have been
classified as broad-line active galactic nuclei (BLAGN); this class includes
QSOs and type 1 Seyferts. Galaxies which only show emission lines with 
FWHM $<$~1000~km~s$^{-1}$ have been classified as narrow emission line 
galaxies (NELGs). 
Galaxies in which we do not detect any emission lines have
been classified simply as `galaxies'. Finally, one object has the spectrum of
a Galactic star.
%although some narrow-line Seyfert 1 galaxies might
%be classified as NELGs with this simple classification scheme, none of the
%NELGs in this sample have line ratios which would classify them as 
%narrow line Seyfert 1s \citep{osterbrock85}.
% what are the sub types of these?

As can be seen in Fig. \ref{fig:optspectra} and Table \ref{tab:optids}, the
sources all have $z<0.8$, and
majority (31/42) of the sources are
classified as NELGs. In all the NELGs we
observe at least one of 
H$\alpha$, [O\,III]\,4954\,\AA/5007\,\AA\ or [O\,II]\,3727\,\AA.
The three objects classified as galaxies have $0.3<z<0.4$, similar to the mean
redshift for the NELGs, $\langle z \rangle = 0.29\pm0.03$, 
so their galaxy classification
is not due to these emission lines being redshifted outside the 
observed spectral
window. The 7 BLAGN, however, have a higher mean redshift 
 than either the NELGs or galaxies, $\langle z \rangle = 0.53\pm0.08$.
More than half of the BLAGN
are identified by a broad Mg\,II\,2798\,\AA\ line. 

Two sources (XMM~J$122656.53$$+$$013125.2$ and 
XMM~J$140057.10$$-$$110120.6$) have emission lines with 
1000~km~s~$<$~FWHM~$<$~2000~km~s$^{-1}$, and therefore might conceivably be
narrow line Seyfert 1 galaxies \citep{osterbrock85}. In
XMM~J$122656.53$$+$$013125.2$ the Mg~II line has a measured FWHM of
1600~km~s$^{-1}$, and the other lines all have
FWHM~$<$~1000~km~s$^{-1}$. However the large ratio of [O~III]/H$\beta$ rules 
out a narrow-line Seyfert 1 galaxy \citep{shuder81}, and suggests
instead that the broad line region is heavily reddened. The Mg~II profile 
could be a composite of emission from the narrow line region and a reddened
line from the broad line region. For XMM~J$140057.10$$-$$110120.6$ only one
emission line is detected (H$\alpha$) with a measured FWHM of 1900~km~s$^{-1}$.
The red, galaxy-dominated continuum leads us to suspect that the nucleus and
broad line region are also reddened in this object, and that the H$\alpha$
line profile is likely to be a composite of emission from the narrow line
region and a reddened broad line.

\subsection{Basic X-ray spectral modeling}
\label{sec:basicfits}

\begin{figure*}
\begin{center}
\leavevmode
\psfig{figure=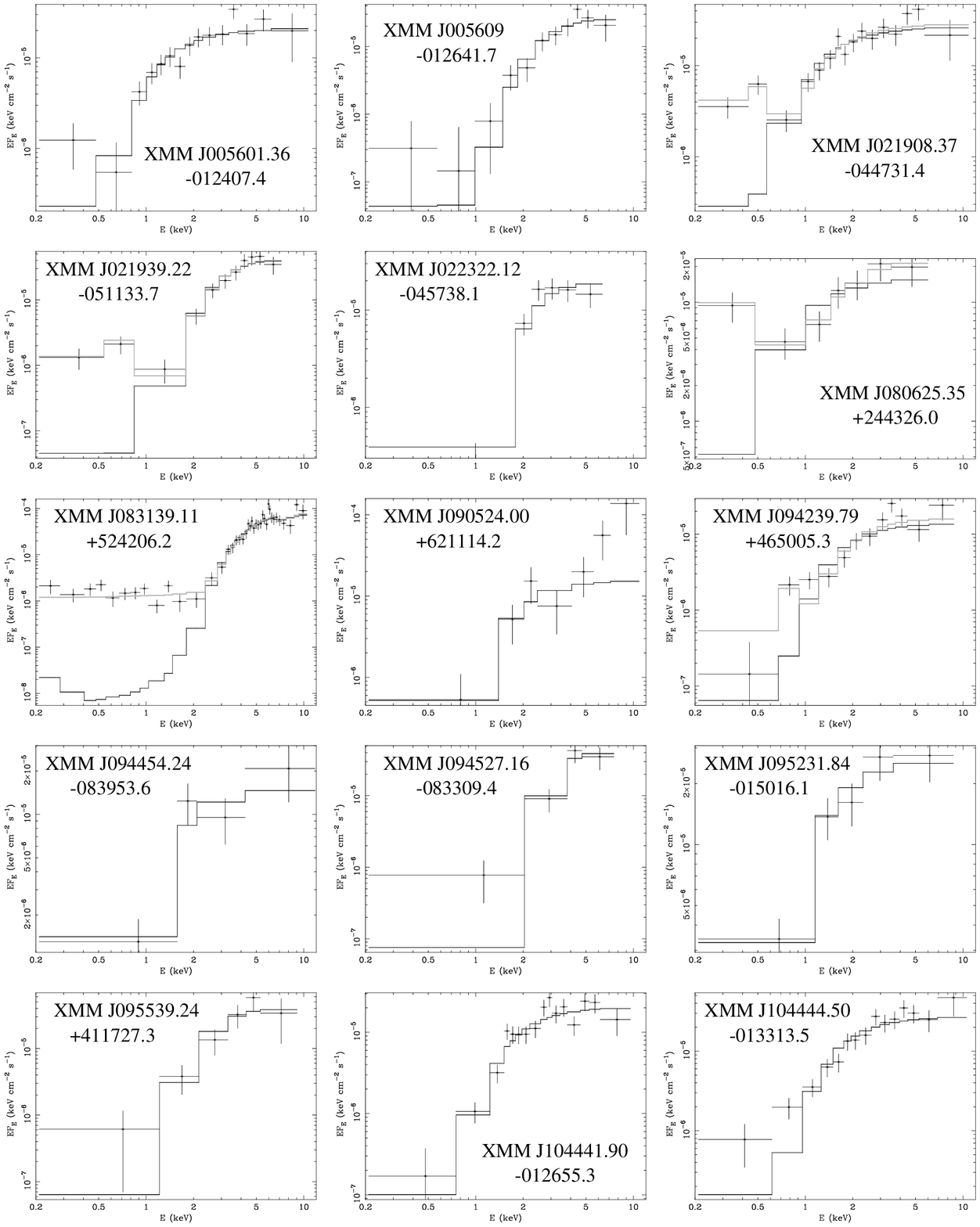,width=160truemm}
\caption{X-ray spectra of the sources (datapoints) together with absorbed
$\Gamma=1.9$ power
law models (black stepped lines). Both model and data have been divided by the
effective area as a function of energy, and the spectra are displayed as 
$E\times F_{E}$ so
that an unabsorbed power law with $\Gamma=2$ would be a horizontal line.
When the absorbed $\Gamma=1.9$ 
power law model does
not provide an acceptable fit to the data, the best alternative model is shown
in grey. For XMM~J$021908.37$$-$$044731.4$, XMM~J$021939.22$$-$$051133.7$, 
and XMM~J$094239.79$$+$$465005.3$ the grey model includes a mekal component,
for XMM~J$083139.11$$+$$524206.2$, XMM~J$133026.09$$+$$241356.7$,
XMM~J$150339.60$$+$$101605.6$ and XMM~J$161736.21$$+$$122901.5$ the grey model
includes an unabsorbed power law component, for XMM~J$080625.35$$+$$244326.0$
the grey model has an ionised absorber, and for XMM~J$122656.53$$+$$013125.2$
the grey model has a power law index of $\Gamma=1.5$.
}
\label{fig:specfits}
\end{center}
\end{figure*}

The X-ray spectra are shown in Fig. \ref{fig:specfits}.
As a starting point for the spectral modeling we considered an absorbed power
law model, corresponding to an AGN viewed through a screen of cold
material. For each object we included a zero-redshift absorber with $\nh$ 
fixed at the Galactic value found from 21\,cm observations 
\citep{dickey90} and a cold absorber at the redshift of the source 
with $\nh$ as a free parameter in the fit.
In fits of this type, particularly in low--moderate signal/noise
data, the power-law photon index $\Gamma$ and absorbing column density $\nh$ 
are somewhat degenerate,
and this can lead to large, highly coupled uncertainties on both
parameters. Therefore we began our spectral fitting with $\Gamma$ fixed at 
a value of 1.9, which is the mean slope found for
unobscured BLAGN in \xmm\ surveys 
\citep{mainieri02,piconcelli02,caccianiga04,mateos05a,page06b}. 

\begin{table*}
\begin{center}
\caption{Absorbed power law fits to the \xmm\ spectra with fixed photon index
$\Gamma = 1.9$. The column labelled `Counts min/bin' indicates the minimum
number of counts per bin used when constructing the spectrum.
$A$ is the power law normalisation in units of 
$10^{-5}$~photons~cm$^{-2}$~s$^{-1}$~keV$^{-1}$.
The `Prob' column gives
the null hypothesis probability corresponding to $\chi^{2}/\nu$. Confidence
limits are given at 95\% for one interesting parameter ($\Delta \chi^{2}=4$); 
confidence
limits that are truncated by the allowed fit range of the
parameter, rather than by $\Delta \chi^{2}=4$ are labelled with `$*$'.
The final column gives the 95\% upper limit on the ratio of a second,
 unabsorbed power law to the primary absorbed power law for those sources which
are well fitted by the single absorbed power law model, as described in Section
\ref{sec:complexspectra}.}
%\input{nhfitstab.tex}
 % Serrendipitous hard xmm sources
 % Table produced automatically by xmmhardpics.for
 \begin{tabular}{lcccccc}
 Source&Counts&$A$&log $N_{H}$&$\chi^{2}/\nu$&Prob&95\% upper limit\\
 &min/bin& &(cm$^{-2}$)&&&$A_{2}/A_{1}$\\
 &&&&&\\
XMM~J$005601.36$$-$$012407.4$&30&$1.7^{+0.4}_{-0.3}$&$22.1^{+0.2}_{-0.2}$& 13 / 13&0.43                 &  0.12 \\%  $XX_01941_00027$
XMM~J$005609.41$$-$$012641.7$&20&$2.3^{+0.9}_{-0.8}$&$22.7^{+0.2}_{-0.2}$&  5 /  9&0.85                 &  0.05 \\%  $XX_01941_00028$
XMM~J$021908.37$$-$$044731.4$&20&$2.2^{+0.4}_{-0.4}$&$22.0^{+0.1}_{-0.2}$& 38 / 14&$6.1\times 10^{-4}$  &   --  \\%  $XX_00693_00176$
XMM~J$021939.22$$-$$051133.7$&20&$3.7^{+1.1}_{-1.0}$&$22.8^{+0.2}_{-0.2}$& 23 /  9&$7.1\times 10^{-3}$  &   --  \\%  $XX_00693_00012$
XMM~J$022322.12$$-$$045738.1$&20&$1.9^{+0.6}_{-0.5}$&$23.1^{+0.2}_{-0.2}$&  4 /  4&0.39                 &  0.02 \\%  $XX_21224_00023$
XMM~J$080625.35$$+$$244326.0$&15&$1.4^{+0.8}_{-0.8}$&$21.9^{+0.5}_{-21.9*}$& 16 /  5&$7.2\times 10^{-3}$&   --  \\%  $XX_02907_00021$
XMM~J$083139.11$$+$$524206.2$&20&$7.0^{+1.3}_{-1.1}$&$23.3^{+0.1}_{-0.1}$&149 / 38&$6.7\times 10^{-15}$ &   --  \\%  $XX_02752_00037$
XMM~J$090524.00$$+$$621114.2$& 5&$1.3^{+1.2}_{-0.8}$&$22.3^{+0.6}_{-0.5}$&  7 /  5&0.25                 &  0.16 \\%  $XX_21387_00038$
XMM~J$094239.79$$+$$465005.3$&20&$1.2^{+0.4}_{-0.4}$&$22.3^{+0.3}_{-0.3}$& 27 / 10&$2.3\times 10^{-3}$  &   --  \\%  $XX_01118_00023$
XMM~J$094454.24$$-$$083953.6$&10&$1.3^{+0.7}_{-0.6}$&$22.3^{+0.4}_{-0.4}$&  2 /  2&0.33                 &  0.22 \\%  $XX_03041_00030$
XMM~J$094527.16$$-$$083309.4$&10&$4.8^{+6.6}_{-3.4}$&$23.3^{+0.4}_{-0.7}$&  3 /  2&0.23                 &  0.07 \\%  $XX_03041_00069$
XMM~J$095231.84$$-$$015016.1$&20&$2.2^{+0.7}_{-0.6}$&$21.9^{+0.2}_{-0.3}$&  1 /  3&0.76                 &  0.26 \\%  $XX_03411_00023$
XMM~J$095539.24$$+$$411727.3$&10&$3.6^{+3.7}_{-2.0}$&$22.8^{+0.5}_{-0.4}$&  3 /  4&0.56                 &  0.08 \\%  $XX_00929_00047$
XMM~J$104441.90$$-$$012655.3$&20&$1.7^{+0.3}_{-0.3}$&$22.3^{+0.1}_{-0.1}$& 17 / 14&0.28                 &  0.03 \\%  $XX_21391_00070$
XMM~J$104444.50$$-$$013313.5$&20&$2.2^{+0.6}_{-0.5}$&$22.0^{+0.2}_{-0.2}$& 18 / 13&0.17                 &  0.13 \\%  $XX_21391_00022$
XMM~J$105014.58$$+$$325041.5$&20&$1.7^{+0.5}_{-0.5}$&$22.3^{+0.2}_{-0.2}$&  1 /  4&0.95                 &  0.07 \\%  $XX_02166_00009$
XMM~J$105743.59$$-$$033402.4$&20&$1.1^{+0.6}_{-0.4}$&$22.5^{+0.3}_{-0.3}$& 12 /  4&$1.6\times 10^{-2}$  &  0.44 \\%  $XX_01035_00100$
XMM~J$111825.42$$+$$073448.3$&20&$1.2^{+0.4}_{-0.3}$&$22.6^{+0.2}_{-0.2}$& 22 / 13&$6.0\times 10^{-2}$  &  0.14 \\%  $XX_02429_00022$
XMM~J$115233.75$$+$$371156.3$& 5&$5.1^{+4.4}_{-3.2}$&$22.4^{+0.5}_{-0.5}$&  1 /  2&0.75                 &  0.27 \\%  $XX_00991_00051$
XMM~J$121312.04$$+$$024929.8$&20&$1.3^{+0.7}_{-0.5}$&$22.5^{+0.3}_{-0.3}$& 10 /  4&$3.7\times 10^{-2}$  &  0.16 \\%  $XX_00377_00092$
XMM~J$122656.53$$+$$013125.2$&20&$5.8^{+0.9}_{-0.8}$&$22.4^{+0.1}_{-0.1}$& 37 / 17&$5.6\times 10^{-3}$  &   --  \\%  $XX_00916_00046$
XMM~J$125701.68$$+$$281230.2$&20&$3.4^{+1.0}_{-0.9}$&$22.8^{+0.2}_{-0.2}$& 17 / 11&0.11                 &  0.05 \\%  $XX_21408_00086$
XMM~J$130129.08$$+$$274037.1$&20&$1.0^{+0.4}_{-0.3}$&$22.7^{+0.2}_{-0.2}$&  5 /  5&0.44                 &  0.06 \\%  $XX_21420_00024$
XMM~J$130458.98$$+$$175451.8$&20&$2.1^{+0.8}_{-0.7}$&$22.6^{+0.2}_{-0.2}$& 12 /  9&0.21                 &  0.03 \\%  $XX_20605_00032$
XMM~J$130529.89$$-$$102141.5$& 5&$1.1^{+1.0}_{-0.6}$&$22.7^{+0.5}_{-0.4}$&  3 /  5&0.69                 &  0.08 \\%  $XX_00379_00095$
XMM~J$133026.09$$+$$241356.7$&20&$2.8^{+0.5}_{-0.4}$&$22.5^{+0.1}_{-0.1}$&117 / 34&$5.6\times 10^{-11}$ &   --  \\%  $XX_00116_00083$
XMM~J$133623.04$$+$$241738.6$&15&$3.6^{+2.1}_{-1.8}$&$22.4^{+0.3}_{-0.5}$&  7 /  3&$8.4\times 10^{-2}$  &  0.24 \\%  $XX_00760_00028$
XMM~J$133913.92$$-$$314421.9$&20&$1.6^{+0.4}_{-0.4}$&$22.3^{+0.2}_{-0.2}$& 10 /  9&0.38                 &  0.13 \\%  $XX_02073_00031$
XMM~J$134656.75$$+$$580316.5$&20&$6.1^{+1.9}_{-1.5}$&$22.9^{+0.2}_{-0.2}$& 20 / 11&$4.5\times 10^{-2}$  &  0.02 \\%  $XX_01182_00009$
XMM~J$140057.10$$-$$110120.6$&20&$1.8^{+0.5}_{-0.5}$&$22.4^{+0.2}_{-0.2}$& 11 / 11&0.46                 &  0.05 \\%  $XX_01142_00152$
XMM~J$140145.11$$+$$025334.6$&10&$1.4^{+0.7}_{-0.6}$&$22.0^{+0.4}_{-0.5}$&  3 /  6&0.81                 &  0.38 \\%  $XX_01046_00117$
XMM~J$140149.53$$-$$111647.8$&20&$1.3^{+0.3}_{-0.3}$&$22.4^{+0.2}_{-0.2}$& 12 / 13&0.51                 &  0.07 \\%  $XX_01142_00015$
XMM~J$141634.78$$+$$113329.2$&20&$2.1^{+0.4}_{-0.4}$&$22.3^{+0.1}_{-0.1}$& 23 / 17&0.16                 &  0.18 \\%  $XX_01183_00081$
XMM~J$145048.99$$+$$191431.2$&20&$1.4^{+0.3}_{-0.3}$&$22.6^{+0.1}_{-0.1}$&  7 / 10&0.72                 &  0.11 \\%  $XX_20252_00072$
XMM~J$150339.60$$+$$101605.6$&20&$7.1^{+2.1}_{-1.7}$&$22.8^{+0.2}_{-0.1}$& 47 / 13&$1.0\times 10^{-5}$  &   --  \\%  $XX_02347_00005$
XMM~J$150558.42$$+$$014104.2$&25&$3.1^{+1.0}_{-0.9}$&$22.4^{+0.2}_{-0.2}$&  7 /  5&0.23                 &  0.06 \\%  $XX_01988_00074$
XMM~J$151650.30$$+$$070907.3$&20&$0.8^{+0.3}_{-0.2}$&$22.3^{+0.3}_{-0.3}$& 11 / 12&0.53                 &  0.22 \\%  $XX_01143_00109$
XMM~J$152146.64$$+$$073113.9$&20&$3.9^{+0.6}_{-0.6}$&$22.3^{+0.1}_{-0.1}$& 29 / 29&0.48                 &  0.07 \\%  $XX_01144_00025$
XMM~J$152254.63$$+$$074534.5$&20&$0.8^{+0.9}_{-0.5}$&$22.8^{+0.6}_{-0.5}$&  4 /  2&0.15                 &  0.05 \\%  $XX_01144_00136$
XMM~J$161646.54$$+$$122015.1$&20&$1.9^{+0.4}_{-0.4}$&$22.2^{+0.2}_{-0.1}$& 12 / 10&0.32                 &  0.17 \\%  $XX_21437_00073$
XMM~J$161736.21$$+$$122901.5$&20&$1.7^{+0.7}_{-0.5}$&$22.0^{+0.3}_{-0.3}$& 29 / 10&$1.3\times 10^{-3}$  &   --  \\%  $XX_21440_00014$
XMM~J$161818.95$$+$$124110.3$&20&$3.5^{+1.8}_{-1.5}$&$22.8^{+0.3}_{-0.3}$&  4 /  2&0.12                 &  0.15 \\%  $XX_21440_00041$
 \end{tabular}
\label{tab:nhfits}
\end{center}
\end{table*}

The results of these fits are given in Table \ref{tab:nhfits}; the
uncertainties listed in Table \ref{tab:nhfits} and all subsequent tables are
95\% for one interesting parameter ($\Delta \chi^{2} = 4$). We consider 
a null hypothesis probability of 1\% to be an appropriate threshold for 
rejection of the spectral model, because at this level, and for our sample 
size, we expect less than 
one rejection by chance if the form of the model is correct. Of the 42 
sources, 33 are therefore acceptably fitted with this model, while 9 
objects ($\sim 20\%$) are not acceptably fitted. In 7 of the 9 cases, the 
poor $\chi^{2}$ is due to an excess of flux in the lowest energy channels 
with respect to the model (see Fig. \ref{fig:specfits}). 

\subsection{Modelling the 9 sources with complex X-ray spectra}
\label{sec:complexspectra}

We now examine in more detail the X-ray spectra of the 9 sources which were 
not well 
fitted 
with a
$\Gamma=1.9$ power law and neutral absorption.
Before considering spectral models with an 
additional soft X-ray component, we investigate the possibility that the poor 
$\chi^{2}$ could be a consequence of our decision to fix $\Gamma=1.9$
in the power law component. There is certainly a range of photon indices 
in the unabsorbed AGN population \citep{mateos05a,mateos05b,page06b}, 
and the fit with 
fixed $\Gamma$ could result in significant residuals for AGN 
which are outliers in this distribution. Therefore,
for the nine objects which had unacceptable $\chi^{2}/\nu$, we refit the 
spectra, this time 
allowing the photon index of the power law to vary as a fit parameter. 
The results of these fits are shown in Table \ref{tab:powerfits}. Four 
of the objects are acceptably fitted with this model, but only one of these,
XMM~J$122656.53$$+$$013125.2$, has a value of $\Gamma$ consistent with a `normal' AGN
spectrum. The other 3 objects (XMM~J$161736.21$$+$$122901.5$, XMM~J$094239.79$$+$$465005.3$, and 
XMM~J$080625.35$$+$$244326.0$) for which this fit yields a reasonable $\chi^{2}/\nu$
have best fit photon indices which are exceedingly hard, $\Gamma < 1.2$. 
Such photon indices are well outside the range of photon indices normally 
observed in unabsorbed AGN 
\citep[$>3.5\sigma$ outliers from the distribution,][]{mateos05a}, 
and therefore we 
consider that the variation in AGN photon indices only provides a 
plausible explanation for the poor fit of the absorbed $\Gamma=1.9$ power law
model to the X-ray spectrum of 
XMM~J$122656.53$$+$$013125.2$.

For the 9 objects not well fitted in Section \ref{sec:basicfits}, we now 
consider a model of photoelectric absorption in which the absorber 
is ionised, rather than
cold, as ionised absorption is often observed in nearby Seyfert 1 galaxies 
\citep[e.g.][]{reynolds97,george98,blustin05}. For this we use 
the `absori' model in {\small XSPEC}, in which the ionization state of the
absorber is defined by the ionization parameter $\xi=L/nr^{2}$, 
where $L$ is the
ionising luminosity of the source, $n$ is the number density of the absorber 
and $r$ is the distance of the absorber from the ionising radiation. The photon
index of the power law was fixed at $\Gamma=1.9$, and the temperature of the
absorber was fixed at 3$\times 10^{4}$K (the fits are not sensitive
to this parameter). The results 
are given in Table \ref{tab:absorifits}. This model produces acceptable 
$\chi^{2}/\nu$ for 4/9 of the sources, but does not give acceptable fits 
for the other 5 objects.

Next we considered two different models which contain an additional component
which is bright at the soft X-ray end of the spectrum. To minimise the number
of free parameters in the fit, we maintain a fixed $\Gamma=1.9$ for the
primary power law component in these fits.

In the first model, we include a second power law with $\Gamma=1.9$, 
but which is not attenuated by 
the column of cold gas intrinsic to the object, and represents a component of 
the primary emission which is scattered into our line of sight, bypassing 
the absorber. The photon indices of both power laws were frozen at a value of 
1.9 in this model. Although such a model is lacking in experimental 
verification from observations of nearby AGN, it has a long heritage 
\citep[e.g.][]{holt80,turner97,franceschini03,caccianiga04}, 
and its functional 
form provides a simple and convenient parameterisation of the relative 
contributions of the primary, absorbed component and the component which 
provides the excess flux at soft energies. 
The results of fitting this model are 
listed in Table \ref{tab:popofits}. All but 2 objects are fitted acceptably  
using this model. We have also used this model to determine 95\% 
upper limits for
the contribution of any soft component in the sources which are acceptably
fitted with a simple absorbed power law model, by increasing the contribution
of the unabsorbed power law until $\Delta \chi^{2}$=4 with respect to the model
without the unabsorbed power law; these upper limits are included in
Table~\ref{tab:nhfits}.

\begin{table*}
\begin{center}
\caption{Absorbed power law fits to the \xmm\ spectra with photon index
$\Gamma$ free to vary. $A$ is the power law normalisation in units of 
$10^{-5}$~photons~cm$^{-2}$~s$^{-1}$~keV$^{-1}$.
The `Prob' column gives
the null hypothesis probability corresponding to $\chi^{2}/\nu$.
Confidence
limits are given at 95\% for one interesting parameter ($\Delta \chi^{2}=4$); 
confidence
limits that are truncated by the allowed fit range of the
parameter, rather than by $\Delta \chi^{2}=4$ are labelled with `$*$'.
}
%\input{powerfitstab.tex}
 % Serrendipitous hard xmm sources
 % Table produced automatically by xmmhardpics.for
 \begin{tabular}{lccccc}
 Source&$\Gamma$&$A$&log $N_{H}$&$\chi^{2}/\nu$&Prob\\
 &&&(cm$^{-2}$)&&\\
 &&&&&\\
XMM~J$021908.37$$-$$044731.4$&$1.5^{+0.5}_{-0.6}$&$1.3^{+1.1}_{-0.8}$&$21.7^{+0.3}_{-21.7*}$& 35 / 13&$1.0\times 10^{-3}$\\
XMM~J$021939.22$$-$$051133.7$&$1.4^{+1.3}_{-0.9*}$&$1.7^{+11.7}_{-1.3}$&$22.7^{+0.3}_{-0.4}$& 22 /  8&$5.1\times 10^{-3}$\\
XMM~J$080625.35$$+$$244326.0$&$1.1^{+0.5}_{-0.3}$&$0.64^{+0.32}_{-0.15}$&$0.00^{+21.5}_{-0.0*}$&  8 /  4&0.10\\
XMM~J$083139.11$$+$$524206.2$&$1.7^{+0.8}_{-0.7}$&$4.3^{+15.9}_{-3.2}$&$23.2^{+0.2}_{-0.2}$&148 / 37&$3.9\times 10^{-15}$\\
XMM~J$094239.79$$+$$465005.3$&$1.0^{+0.4}_{-0.3}$&$0.35^{+0.26}_{-0.13}$&$21.5^{+0.5}_{-0.6}$& 17 /  9&$5.8\times 10^{-2}$\\
XMM~J$122656.53$$+$$013125.2$&$1.5^{+0.3}_{-0.4}$&$3.6^{+1.8}_{-1.2}$&$22.3^{+0.2}_{-0.2}$& 31 / 16&$1.4\times 10^{-2}$\\
XMM~J$133026.09$$+$$241356.7$&$0.5^{+0.2}_{-0.0*}$&$0.31^{+0.05}_{-0.04}$&$0.0^{+20.8}_{-0.0*}$& 74 / 33&$6.0\times 10^{-5}$\\
XMM~J$150339.60$$+$$101605.6$&$0.5^{+0.5}_{-0.0*}$&$0.55^{+0.78}_{-0.09}$&$21.8^{+0.6}_{-0.4}$& 35 / 12&$4.6\times 10^{-4}$\\
XMM~J$161736.21$$+$$122901.5$&$0.5^{+0.4}_{-0.0*}$&$0.29^{+0.23}_{-0.04}$&$0.0^{+21.5}_{-0.0*}$&  5 /  9&0.85\\
 \end{tabular}
\label{tab:powerfits}
\end{center}
\end{table*}

\begin{table*}
\begin{center}
\caption{Fits to the \xmm\ spectra with a power law of
$\Gamma=1.9$ passing through an ionized medium. $A$ is the 
power law normalisation
in units of $10^{-5}$~photons~cm$^{-2}$~s$^{-1}$~keV$^{-1}$. The ionisation
parameter $\xi$ is in units of erg~cm~s$^{-1}$. Confidence
limits are given at 95\% for one interesting parameter ($\Delta \chi^{2}=4$);  
confidence
limits that are truncated by the allowed fit range of the
parameter, rather than by $\Delta \chi^{2}=4$ are labelled with `$*$'.
The `Prob' column gives
the null hypothesis probability corresponding to $\chi^{2}/\nu$.}
%\input{absorifitstab.tex}
 % Serrendipitous hard xmm sources
 % Table produced automatically by xmmhardpics.for
 \begin{tabular}{lccccc}
 Source&$A$&log $N_{H}$&log $\xi$&$\chi^{2}/\nu$&Prob\\
 &&(cm$^{-2}$)&&&\\
 &&&&&\\
XMM~J$021908.37$$-$$044731.4$&$2.4^{+0.4}_{-0.4}$&$22.2^{+0.1}_{-0.2}$&$ 0.8^{+0.3}_{-0.5}$& 11 / 13&0.59\\
XMM~J$021939.22$$-$$051133.7$&$3.8^{+1.1}_{-1.0}$&$23.0^{+0.2}_{-0.2}$&$ 1.4^{+0.2}_{-0.2}$& 13 /  8&0.11\\
XMM~J$080625.35$$+$$244326.0$&$2.0^{+0.8}_{-0.7}$&$22.9^{+0.4}_{-0.5}$&$ 2.4^{+0.5}_{-0.7}$&  1 /  4&0.97\\
XMM~J$083139.11$$+$$524206.2$&$6.7^{+1.2}_{-1.1}$&$23.4^{+0.1}_{-0.1}$&$ 1.6^{+0.1}_{-0.1}$&128 / 37&$6.6\times 10^{-12}$\\
XMM~J$094239.79$$+$$465005.3$&$1.2^{+0.4}_{-0.4}$&$22.4^{+0.2}_{-0.3}$&$ 0.2^{+0.8}_{-2.2*}$& 25 /  9&$2.7\times 10^{-3}$\\
XMM~J$122656.53$$+$$013125.2$&$5.8^{+1.0}_{-0.8}$&$22.5^{+0.2}_{-0.1}$&$-1.2^{+2.4}_{-0.8}$& 37 / 16&$2.4\times 10^{-3}$\\
XMM~J$133026.09$$+$$241356.7$&$2.7^{+0.4}_{-0.4}$&$22.6^{+0.1}_{-0.1}$&$ 1.0^{+0.1}_{-0.2}$& 82 / 33&$4.9\times 10^{-6}$\\
XMM~J$150339.60$$+$$101605.6$&$7.3^{+2.0}_{-1.7}$&$23.0^{+0.1}_{-0.2}$&$ 1.3^{+0.2}_{-0.3}$& 40 / 12&$8.8\times 10^{-5}$\\
XMM~J$161736.21$$+$$122901.5$&$1.9^{+0.6}_{-0.5}$&$22.3^{+0.2}_{-0.2}$&$ 0.6^{+0.5}_{-0.9}$& 19 /  9&$2.8\times 10^{-2}$\\
 \end{tabular}
\label{tab:absorifits}
\end{center}
\end{table*}

\begin{table*}
\begin{center}
\caption{Two component fits to the \xmm\ spectra consisting of two power laws,
of which one is absorbed by cold material, and the other is not. $A_{1}$
is the normalisation of the absorbed power law and $A_{2}$ is the
normalisation of the unabsorbed component, in units of 
$10^{-5}$~photons~cm$^{-2}$~s$^{-1}$~keV$^{-1}$. Both power laws have fixed
$\Gamma=1.9$. Confidence
limits are given at 95\% for one interesting parameter ($\Delta \chi^{2}=4$); 
confidence
limits that are truncated by the allowed fit range of the
parameter, rather than by $\Delta \chi^{2}=4$ are labelled with `$*$'.
The `Prob' column gives the null hypothesis probability
corresponding to $\chi^{2}/\nu$.}  
%\input{popofitstab.tex}
 % Serrendipitous hard xmm sources
 % Table produced automatically by xmmhardpics.for
 \begin{tabular}{lccccc}
 Source&$A_{1}$&$A_{2}$&log $N_{H}$&$\chi^{2}/\nu$&Prob\\
 &&&(cm$^{-2}$)&&\\
 &&&&&\\
XMM~J$021908.37$$-$$044731.4$&$2.2^{+0.5}_{-0.5}$&$0.32^{+0.12}_{-0.13}$&$22.3^{+0.2}_{-0.2}$& 16 / 13&0.27\\
XMM~J$021939.22$$-$$051133.7$&$4.1^{+1.3}_{-1.1}$&$0.11^{+0.05}_{-0.05}$&$22.9^{+0.3}_{-0.3}$&  7 /  8&0.54\\
XMM~J$080625.35$$+$$244326.0$&$1.5^{+1.1}_{-0.9}$&$0.56^{+0.23}_{-0.27}$&$22.7^{+0.6}_{-0.5}$&  4 /  4&0.47\\
XMM~J$083139.11$$+$$524206.2$&$7.6^{+1.4}_{-1.3}$&$0.13^{+0.02}_{-0.03}$&$23.4^{+0.1}_{-0.1}$& 39 / 37&0.38\\
XMM~J$094239.79$$+$$465005.3$&$1.3^{+0.4}_{-0.4}$&$0.05^{+0.05}_{-0.05}$&$22.4^{+0.3}_{-0.3}$& 23 /  9&$5.5\times 10^{-3}$\\
XMM~J$122656.53$$+$$013125.2$&$5.7^{+0.7}_{-0.8}$&$0.00^{+0.09}_{-0.00*}$&$22.4^{+0.1}_{-0.1}$& 37 / 16&$2.0\times 10^{-3}$\\
XMM~J$133026.09$$+$$241356.7$&$3.0^{+0.5}_{-0.4}$&$0.20^{+0.05}_{-0.04}$&$22.6^{+0.1}_{-0.1}$& 27 / 33&0.78\\
XMM~J$150339.60$$+$$101605.6$&$7.9^{+2.2}_{-1.9}$&$0.21^{+0.09}_{-0.08}$&$22.9^{+0.1}_{-0.1}$& 22 / 12&$3.7\times 10^{-2}$\\
XMM~J$161736.21$$+$$122901.5$&$2.4^{+1.1}_{-0.8}$&$0.23^{+0.09}_{-0.10}$&$22.5^{+0.3}_{-0.3}$& 11 /  9&0.30\\
 \end{tabular}
\label{tab:popofits}
\end{center}
\end{table*}

In the second model, we include an optically-thin thermal plasma component 
(`mekal' in {\small XSPEC}) instead of an additional power law. 
We do not include any attenuation of 
this component by the intrinsic absorber, because it is intended to represent 
a soft X-ray line emitting plasma which lies outside the central, obscured 
region of the AGN. Such a component of line emitting plasma has long been 
seen in 
Seyfert 2s \citep[e.g.][]{marshall93,ueno94,turner97,ogle00}. 
Observations with
the gratings onboard \xmm\ and \chandra\ show that the gas is photoionised 
\citep[e.g.][]{sako00,kinkhabwala02,brinkman02} rather than collisionally 
ionised (as in
the mekal model). However, the two cases are indistinguishable at the 
resolution and signal to noise ratio of our data. The fit results are given in 
Table \ref{tab:mekalfits}; this model is quite successful, in that it 
provides an acceptable fit for 6 of the 9 sources.  

\begin{table*}
\begin{center}
\caption{Two component fits to the \xmm\ spectra consisting of an absorbed,
$\Gamma=1.9$ power law, and an optically-thin thermal plasma. $A$ is the 
power law normalisation
in units of $10^{-5}$~photons~cm$^{-2}$~s$^{-1}$~keV$^{-1}$.
The `Prob' column gives the null hypothesis probability
corresponding to $\chi^{2}/\nu$. Confidence
limits are given at 95\% for one interesting parameter ($\Delta \chi^{2}=4$); 
confidence
limits that are truncated by the allowed fit range of the
parameter, rather than by $\Delta \chi^{2}=4$ are labelled with `$*$'.}  
%\input{mekalfitstab.tex}
 % Serrendipitous hard xmm sources
 % Table produced automatically by xmmhardpics.for
 \begin{tabular}{lccccc}
 Source&$A$&log $N_{H}$&kT&$\chi^{2}/\nu$&Prob\\
 &&(cm$^{-2}$)&(keV)&&\\
 &&&&&\\
XMM~J$021908.37$$-$$044731.4$&$2.4^{+0.5}_{-0.5}$&$22.1^{+0.2}_{-0.2}$&$0.23^{+0.07}_{-0.07}$&  10 / 12&0.64\\
XMM~J$021939.22$$-$$051133.7$&$3.8^{+1.1}_{-1.0}$&$22.8^{+0.2}_{-0.2}$&$0.30^{+0.11}_{-0.08}$&  3 /  7&0.88\\
XMM~J$080625.35$$+$$244326.0$&$1.9^{+0.7}_{-0.7}$&$22.4^{+0.2}_{-0.5}$&$0.43^{+0.40}_{-0.23}$&  1 /  3&0.93\\
XMM~J$083139.11$$+$$524206.2$&$7.2^{+1.2}_{-1.2}$&$23.3^{+0.1}_{-0.1}$&$1.00^{+0.00*}_{-0.11}$& 77 / 40&$8.6\times 10^{-5}$\\
XMM~J$094239.79$$+$$465005.3$&$1.4^{+0.4}_{-0.4}$&$22.5^{+0.1}_{-0.2}$&$1.00^{+0.00*}_{-0.18}$& 16 /  8&$4.9\times 10^{-2}$\\
XMM~J$122656.53$$+$$013125.2$&$5.7^{+1.0}_{-0.8}$&$22.4^{+0.1}_{-0.1}$&$1.00^{+0.00*}_{-0.90*}$& 37 / 15&$1.2\times 10^{-3}$\\
XMM~J$133026.09$$+$$241356.7$&$2.9^{+0.4}_{-0.4}$&$22.5^{+0.1}_{-0.1}$&$0.29^{+0.06}_{-0.05}$& 44 / 32&$7.5\times 10^{-2}$\\
XMM~J$150339.60$$+$$101605.6$&$7.3^{+2.3}_{-1.7}$&$22.8^{+0.2}_{-0.1}$&$1.00^{+0.00*}_{-0.21}$& 27 / 11&$4.0\times 10^{-3}$\\
XMM~J$161736.21$$+$$122901.5$&$2.1^{+0.8}_{-0.6}$&$22.3^{+0.3}_{-0.2}$&$0.41^{+0.59}_{-0.20}$& 17 /  8&$2.9\times 10^{-2}$\\
 \end{tabular}
\label{tab:mekalfits}
\end{center}
\end{table*}

\section{Discussion}
\label{sec:discussion}

\subsection{Optical identifications of the hard spectrum sources}
\label{sec:discussion_optical}

The hardness ratio selection has produced a sample with considerably
different optical %emission line 
properties to the general X-ray
population at $S_{0.5-4.5}>10^{-14}$~erg~cm$^{-2}$~s$^{-1}$. 
Of our sample of 42 objects, 31 are identified as NELGs,
compared to $<20$\% of the overall X-ray population at this flux
level \citep{barcons03}. This is as expected from AGN unification schemes, 
because the hardness ratio traces absorption. 
It is in line with the findings of
several {\em Chandra} and {\em XMM-Newton} surveys, that sources with large 
hardness ratios are predominantly narrow line objects 
\citep[e.g.][]{hasinger01,barger02,silverman05} 
and confirms the specific findings of 
\citet{dellaceca04} and \citet{caccianiga04} that a large fraction of \xmm\
sources with $HR2> -0.3$ are narrow line objects. 
In contrast, BLAGN, which are the 
majority source population in simple flux limited surveys of
equivalent depth \citep{barcons03,fiore03}, account for only 7 out of
42 of our sources.  These emission line
demographics are also remarkably different to those found in hard
spectrum sources selected at lower energy: in the hard-spectrum \ros\ 
sample of \citet{page01}, the majority of the sources are BLAGN with
only $\sim 20\%$ identified as NELGs.  There are many
examples of X-ray absorbed AGN in the literature which appear to show far less
attenuation to their broad lines than would be expected from a
standard Galactic mixture of gas and dust \citep[e.g.][]{loaring03,akiyama00}.
However, in most cases the absorbing media
responsible for hardening the X-ray spectra of the sources in our sample
have successfully suppressed the AGN continuum and broad emission
lines.
%comparison with rosat hard source sample

\begin{figure}
\begin{center}
\leavevmode
\psfig{figure=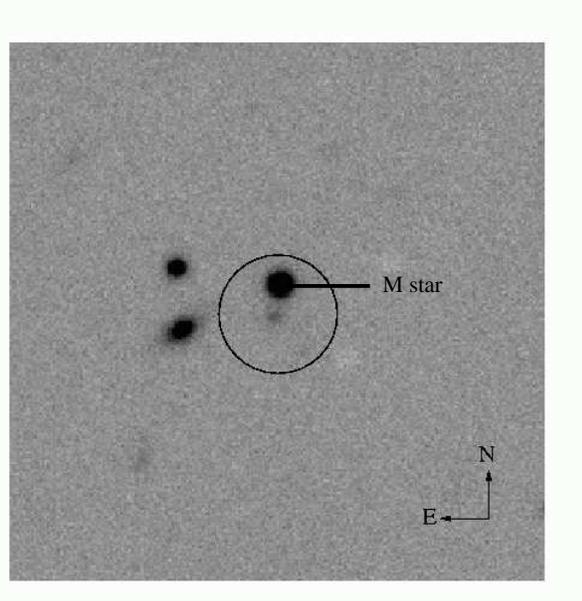,width=80truemm}
\caption{$1\arcmin \times 1\arcmin$ r band image of XMM~J$104444.50-013313.5$.
The black circle is a $6\arcsec$ radius circle centred on the X-ray source
position. The brigher counterpart is the M star for which we have an
optical spectrum. The fainter counterpart is closer to the centre of the circle
and thus to the X-ray source position.}
\label{fig:mstar}
\end{center}
\end{figure}

Only 1 source
(XMM~J$104444.50-013313.5$)
out of 42 has an optical counterpart which is a Galactic star, compared to 11\%
of the overall X-ray population at 
$S_{0.5-4.5}>2\times 10^{-14}$~erg~cm$^{-2}$~s$^{-1}$ \citep{barcons03}. In fact it is suprising
that even one source in our sample should turn out to be a star, because coronally active stars
have optically thin thermal spectra which peak in the range 0.7--1 keV, and
thus they should be excluded from our sample by our hardness ratio selection (Eq. \ref{eq:selection}).
This is well established empirically: almost 
all of the stars in the \xmm\ Bright
Serendipitous Survey have $HR2 < -0.3$ \citep{dellaceca04}. The X-ray spectrum
of XMM~J$104444.50-013313.5$ has a similar form to the rest of our sample, 
and is consistent with an absorbed $\Gamma=1.9$ power law shape. This would be
highly anomalous for a Galactic star \citep[e.g.][]{page06b}. Therefore we must
consider the possibility that the M star is not the correct optical counterpart
to XMM~J$104444.50-013313.5$. In Fig. \ref{fig:mstar} we show a 
$1\arcmin \times 1\arcmin$
region, centred on XMM~J$104444.50-013313.5$, from an 
R band image taken with the Wide Field
Imager on the 2.2m telescope in La Silla. A second optical counterpart
is seen closer to the X-ray source position than the M star, but 2.9 magnitudes
fainter. %+-0.3 mag
We consider that this fainter optical source is more likely to be the
correct optical counterpart than the M star, and that the proximity of the 
M star to the X-ray source is coincidental. If this is correct, then 
XMM~J$104444.50-013313.5$ should be assigned to the optically-faint subsample,
rather than optically-bright subsample.

\begin{figure}
\begin{center}
\leavevmode
\psfig{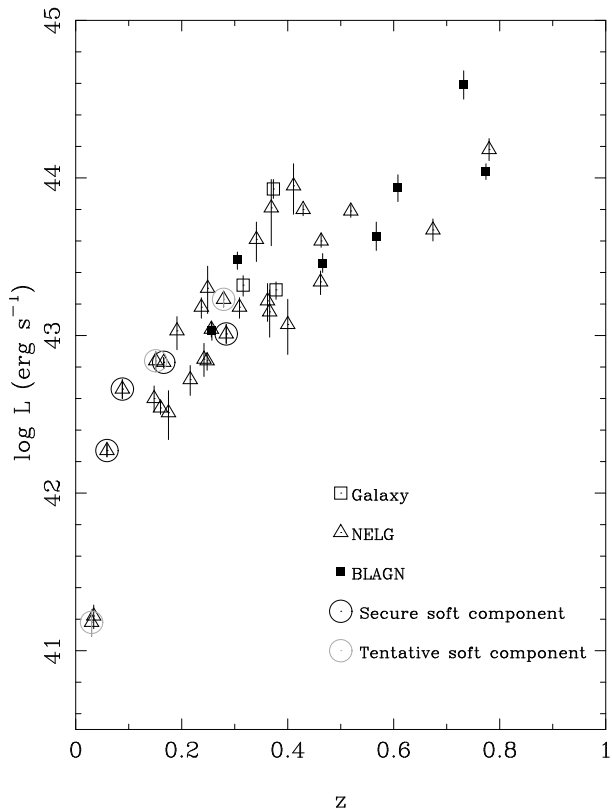}
\caption{2-10 keV luminosities of the sample as a function of redshift. The
luminosities have been computed for the primary power law component in the
best fitting model for each source, and are corrected for
 absorption. The sources enclosed in circles are those which are best 
fit with a soft component in addition to an absorbed power law. The black
circles indicate sources for which the detection of a soft component is 
secure, while the grey
circles indicate sources for which the best-fit model includes a soft 
component, but for which the alternative ionised-absorber model also produces
an acceptable fit (i.e. tentative detections of the soft component). 
Note that the most
luminous source is XMM~J$122656.53$$+$$013125.2$, which has 
an optical spectrum
that is dominated by narrow emission lines, but is classified as a
BLAGN because it has an Mg\,II line with FWHM$> 1000$~km~s$^{-1}$.}
\label{fig:zbest210lumin}
\end{center}
\end{figure}

\subsection{Redshift and luminosity distributions}
\label{sec:discussion_redshifts}

For each source we have calculated
the intrinsic, rest-frame 2-10 keV luminosity, based on the 
unabsorbed primary power law continuum from the best-fit spectral model, shown
in Fig. \ref{fig:specfits}. 
These luminosities are shown as a function of
redshift in Fig. \ref{fig:zbest210lumin}. 
As expected from a flux limited sample,\footnote{A uniform flux limit has not
been imposed on the sample; nonetheless the sample is flux-limited in that the
distance to which an object of any particular luminosity can be detected 
is determined by its flux.}
 luminosity is strongly correlated with
redshift: the linear correlation coefficient of log~$L_{2-10}$ and z is
0.84, and the probability that there is no correlation is $<10^{-11}$. 
The BLAGN of the sample have higher mean redshifts and luminosities
($\langle z\rangle = 0.53\pm0.08$, 
$\langle \log L_{2-10}\rangle = 43.7\pm0.19$) 
than the NELGs 
($\langle z\rangle = 0.30\pm0.03$, 
$\langle \log L_{2-10}\rangle = 43.0\pm0.12$), but the differences are
only of moderate statistical significance: 
according to the Kolmogorov-Smirnov test 
both the redshift 
distributions and the luminosity distributions differ between the 
BLAGN and NELGs with 98\% confidence.

All but 2 of the sources have 2-10 keV luminosities in excess of
$10^{42}$~erg~s$^{-1}$, and therefore undoubtably contain AGN.
The other two objects, which have 2-10 keV 
luminosities of $\sim 10^{41}$~erg~s$^{-1}$, do not display 
 deep Balmer absorption lines in their optical spectra, 
which would indicate a  population of young 
stars. One of these two sources, 
XMM~J$161736.21+122901.5$, has emission lines which
classify it as a Seyfert 2, 
%(checked encarni's thesis)
and hence is unambiguously AGN-dominated.
The other, XMM~J$130458.98+175451.8$, has only a weak [O\,II]\,3727\,\AA\ line,
and no H$\alpha$ emission can be discerned, suggesting that it is not 
strongly star forming, and hence that its X-ray emission must also come from an
obscured AGN. 

Most of the objects in our sample 
have luminosities typical of Seyfert galaxies ($10^{41} <
L_{2-10} < 10^{44}$). However, the three highest redshift objects have 
$L_{2-10} > 10^{44}$~erg~s$^{-1}$, within the luminosity range of 
QSOs. Two of these would be classified as `type 2' QSOs according to their
optical spectra: XMM~J$022322.12-045738.1$ shows
only narrow lines, and XMM~J$122656.53+013125.2$ is dominated by narrow lines
(although the FWHM of Mg\,II exceeds 1000 km\,s$^{-1}$, leading to a
BLAGN classification according to our scheme). The other, 
XMM~J$080625.35+244326.0$, shows little evidence
for optical obscuration, exhibiting a strong ultraviolet continuum and broad
emission lines (Fig. \ref{fig:optspectra}).

All of the sources in our sample have been imaged at 1.4 GHz by the Very Large
Array  surveys FIRST \citep{becker95} or 
NVSS \citep{condon98}. Only two of our
X-ray sources
have radio counterparts with 1.4 GHz flux density $>$ 2.5\,mJy:
XMM~J$150339.60+101605.6$ with a flux density of 113\,mJy and
XMM~J$104444.50-013313.5$ with a flux density of 4.3\,mJy. The position of the 
radio
counterpart to XMM~J$104444.50-013313.5$ is consistent with the faint
counterpart near the centre of the error circle, rather than the M star that we
observed with the WHT.  The small rate of radio detections implies that the
sample is almost completely dominated by radio-quiet objects.

\subsection{X-ray spectral characteristics}
\label{sec:discussion_xrayspecs}

The spectral fits show that the basic hypothesis that the sources have
spectra similar to normal, unobscured AGN, attenuated by large columns
of cold absorbing material, is adequate for most of the sources. The
underlying continua are in almost all cases consistent with a typical AGN 
power law of $\Gamma=1.9$. Only one object appears to require a 
different spectral index: for XMM~J$122656.53$$+$$013125.2$ the only model that 
produces an acceptable fit to the X-ray spectrum has 
$\Gamma=1.5\pm0.3$. In four of the remaining 8 cases for which the
basic $\Gamma=1.9$
power law and cold absorber are rejected, an ionized absorber provides an
acceptable fit to the X-ray spectrum. However, in only one case does the
ionized absorber provide the highest null-hypothesis probability (i.e. best
fit) of all the models tried:
XMM~J$080625.35$$+$$244326.0$, which is also the only BLAGN in the sample 
for which the
simple power law and cold absorber model fails. Its intrinsic luminosity 
($L_{2-10}\sim 10^{44}$~erg~s$^{-1}$) 
%and column density  ($10^{23}$~cm$^{-2}$) 
is similar to that of 
the archetypal warm-absorber Seyfert 1 H$0557-385$ \citep{ashton06}, 
suggesting that this is a
plausible model.
It is interesting to note that
in their study of 41 serendipitous \xmm\ sources, \citet{piconcelli02} found
evidence for an ionized absorber in only one source,
XMMU~J$140127.7$$+$$025603$, which 
is also a BLAGN. Furthermore, in a study of the 13$^{H}$ Deep Field, 
\citet{page06b} found that at least one of the X-ray absorbed BLAGN 
(source 101) has an ionised absorber.
Unfortunately XMM~J$080625.35$$+$$244326.0$ has
one of the poorest signal-to-noise X-ray spectra in our sample, 
and the models incorporating an additional
component of soft X-rays also result in acceptable fits. Nonetheless, the
success of the ionised absorber model for this object is consistent with a
picture in which a significant fraction of X-ray absorbed BLAGN posess 
ionised absorbers.

For the remaining 7 sources, all classified as NELGs, for which the simplest
model (a $\Gamma=1.9$ power law and cold absorber) is rejected, the best fits
are obtained when an additional soft component is included. 
For 4 of these
sources this is the only viable model, so we consider these to be 
`secure' detections
of the soft component. We consider the other 3 sources to be 
`tentative' detections of the soft component, because we cannot rule out the 
alternative model with an 
ionised absorber on $\chi^{2}$ grounds.
Both models for the
soft component are successful for most, but not all, of these 7 objects: the
power law soft excess results in acceptable fits for all but
XMM~J$094239.79$$+$$465005.3$, and the thermal plasma soft excess results in
acceptable fits for all but XMM~J$083139.11$$+$$524206.2$ and
XMM~J$150339.60$$+$$101605.6$. It is somewhat suprising that the power law soft
component is as successful as the thermal plasma, given that the soft X-ray
emission in nearby Seyfert 2s is readily resolved into lines by grating
observations \citep[e.g.][]{kinkhabwala02,sako00,sambruna01}. However, it
should be noted that an unabsorbed power law soft component was required to 
reproduce
the soft X-ray emission observed in NGC~4151 with \asca\ \citep{george98}, even
though this emission was also subsequently resolved into line emission and
recombination continua \citep{ogle00,schurch04}. As our EPIC spectra 
are of much lower signal/noise than the archival \asca\
data on NGC~4151, we cannot rule out a soft component dominated by emission
lines and recombination continua in XMM~J$083139.11$$+$$524206.2$ and
XMM~J$150339.60$$+$$101605.6$.

\begin{figure*}
\begin{center}
\leavevmode
\psfig{figure=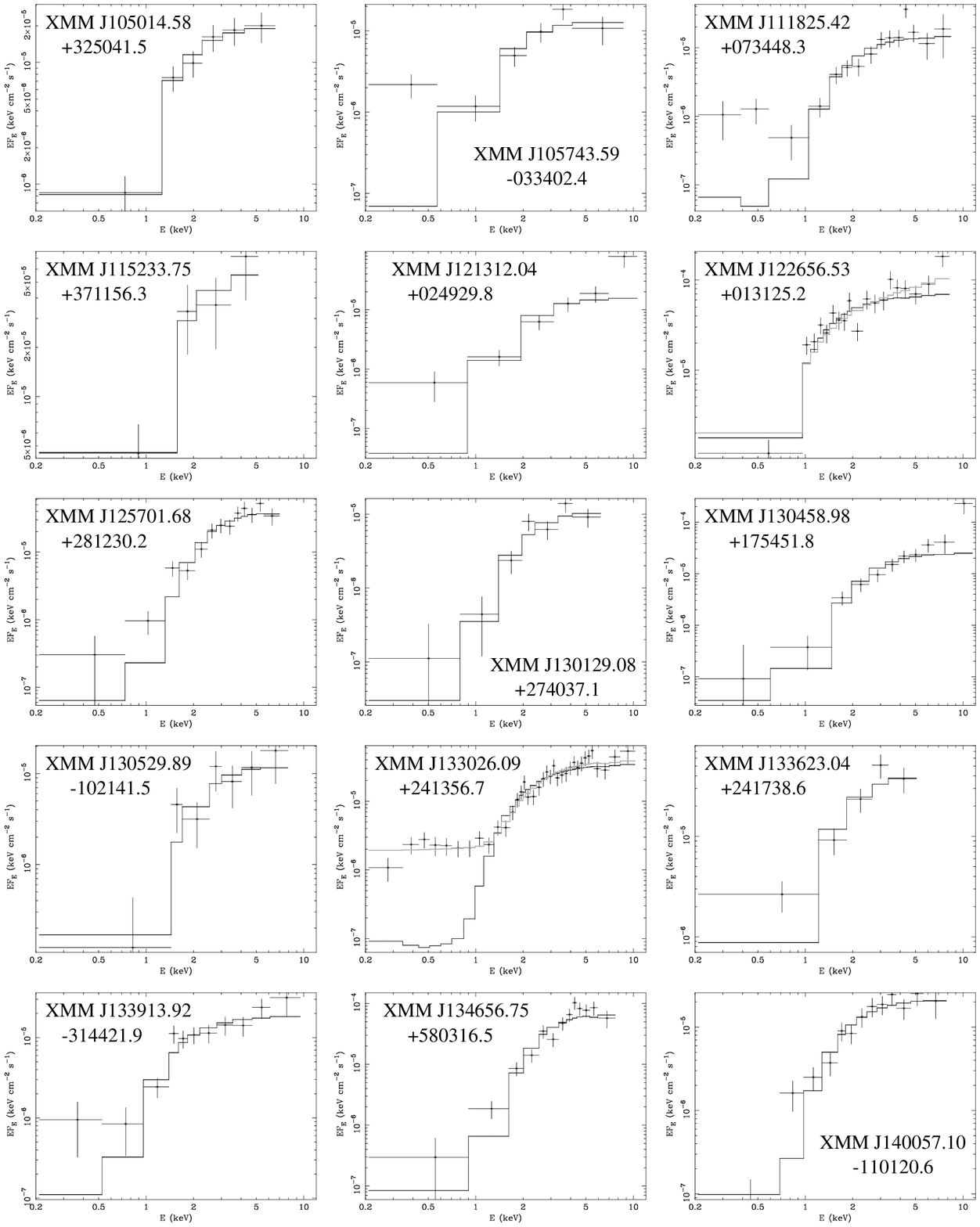,width=160truemm}
\end{center}
{\bf Figure \ref{fig:specfits}} {\it continued}
\end{figure*}
\begin{figure*}
\begin{center}
\leavevmode
\psfig{figure=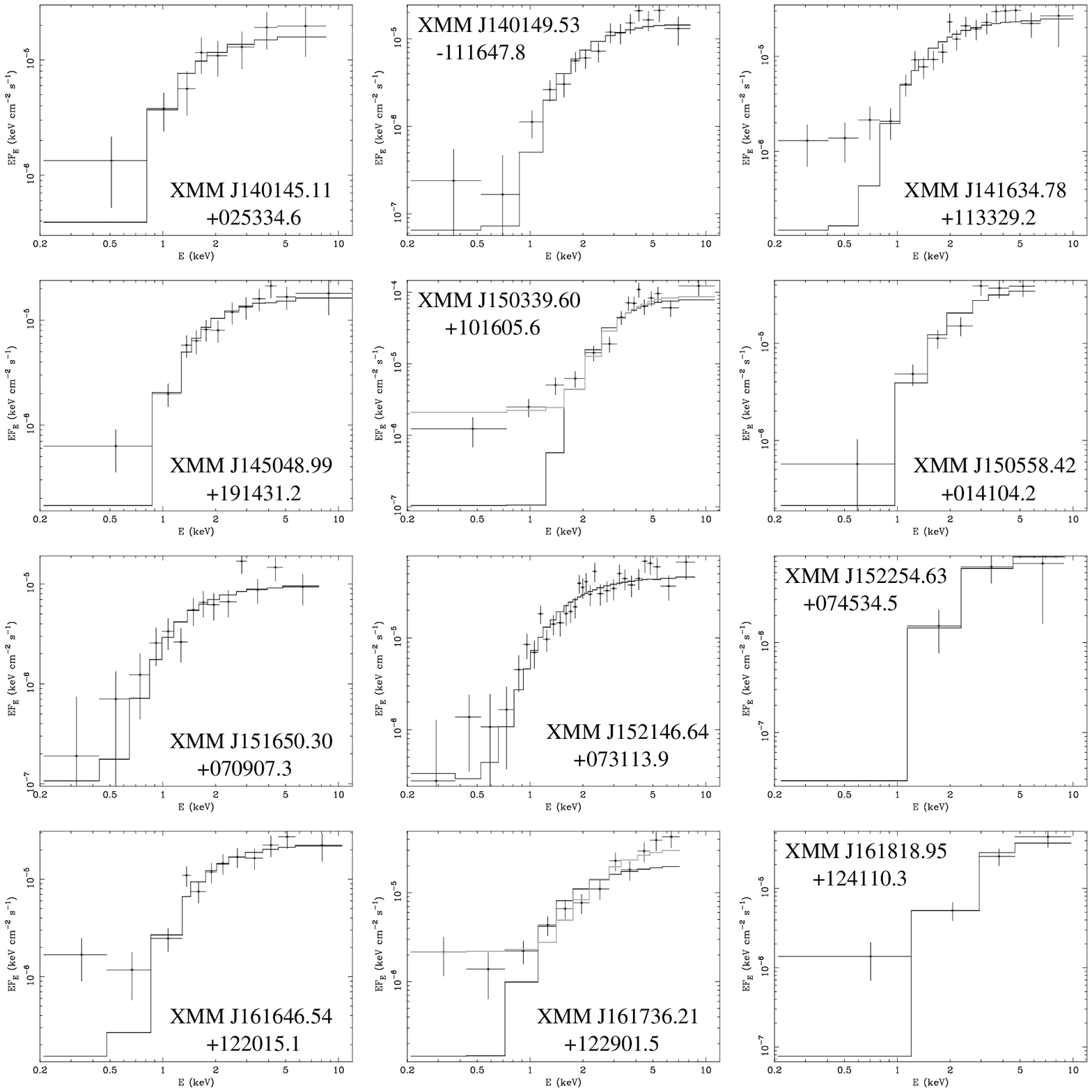,width=160truemm}
\end{center}
{\bf Figure \ref{fig:specfits}} {\it continued}
\end{figure*}

\begin{figure}
\begin{center}
\leavevmode
\psfig{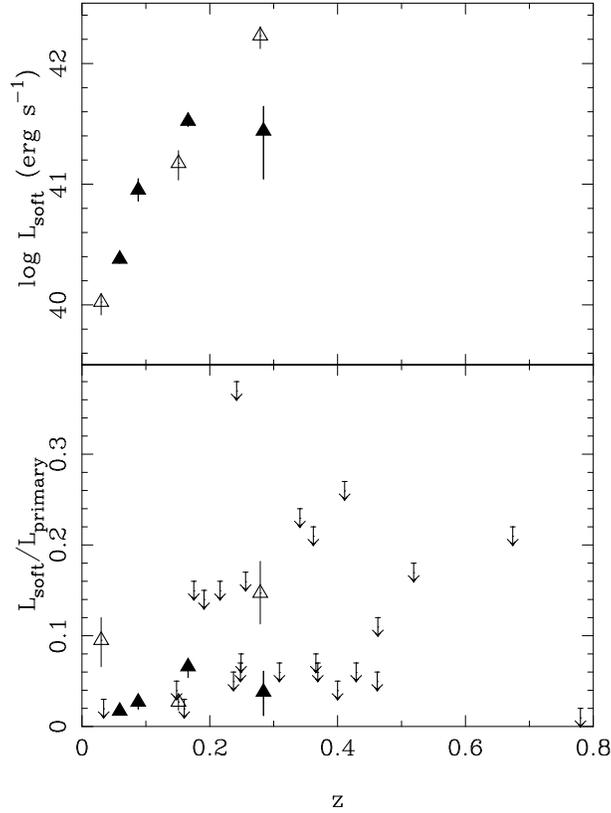}
\caption{Top panel: 0.5-2 keV luminosities of the soft components 
(based on the power law model) for the 7 NELGs with a spectrum more complex
than an absorbed power law. The filled symbols indicate those NELGs for which
acceptable fits were only found with an additional soft component (i.e. solid
detections of the soft component), while the
open symbols indicate those objects for which acceptable fits could also be
found with an ionised absorber instead of a soft component (i.e. tentative
detections of the soft component). Bottom
panel: ratios of the soft (unabsorbed) power law normalisation to that of the
 primary (absorbed) power
law, for the NELGs. Filled and open triangles indicate the same objects as in
the top panel, while upper limits are given for those objects that could be 
fitted with the absorbed
power law model (see Table \ref{tab:nhfits}).}
\label{fig:zsoft52lumin}
\end{center}
\end{figure}

It is notable that the 7 NELGs with tentative or secure detections of a 
soft X-ray component are
toward the lower redshift, lower luminosity end of the sample (see
Fig. \ref{fig:zbest210lumin}). All of them have $z < 0.3$ and $L_{2-10} <
3\times 10^{43}$~erg~s$^{-1}$. This may be a natural consequence of the 
way we have selected and analysed the sample.
Firstly, for higher redshift sources a smaller proportion of
the soft component is visible within the 0.2-10 keV energy range of
EPIC, making it more difficult to detect the soft component. 
%selection effects due to HR selection
%mean log NH=22.53+-0.05
Secondly, the $HR2-\sigma_{HR2}>-0.3$ selection
criterion will exclude objects from the sample if the soft component is too 
strong. For a typical source with $\Gamma=1.9$, $\log \nh=22.5$ and
$\sigma_{HR2}=0.15$, the
strength of soft component that will remove it from the sample declines from 16
per cent of the primary power law at $z=0.3$ to only 7 percent at $z=0.6$.
Thus a soft excess is simultaneously more difficult to detect, and more 
likely to violate our selection criterion, at $z>0.3$.
The upper panel of Fig. \ref{fig:zsoft52lumin} shows the 0.5-2 keV 
luminosities of the soft 
components
in these 7 objects as a function of redshift, based on the model with a
power-law soft component. The soft component luminosities, like the primary
emission, appear to be highly correlated with redshift. This may itself be a
selection effect related to our ability to detect the soft component in the
presence of the primary absorbed power law.  The most powerful soft 
component of the 7 sources has a 0.5-2 keV
luminosity of $\sim 2\times 10^{42}$~erg~s$^{-1}$, but this is a `tentative'
soft component as the spectrum of XMM~J$021908.37-044731.4$ can also be
modelled with an ionised absorber; of the 4 secure detections
of the soft component, the most powerful has a luminosity of 
$\sim 3\times 10^{41}$~erg~s$^{-1}$.
In the lower panel of 
Fig. \ref{fig:zsoft52lumin} we show the ratios of the soft component 
flux to the
intrinsic (i.e. unabsorbed) flux of the primary power law component. We also
show the upper limits for soft component contributions for the NELGs in which
no soft component is detected. All of the secure soft component detections 
have luminosities equivalent to less than 10\% that of the primary power law,
and more than half of the upper limits are also below this level. Only 1
tentatively detected soft component has a luminosity $>$10\% of the primary
power law.

%bit to insert somewhere:
%To quantify this effect, we have calculated what strength of
%soft component would produce $\Delta HR2$ of $-0.3$, which would be sufficient
%to exclude 50 per cent of the sample. For a typical object in our sample, 
%with $\Gamma=1.9$, $z=0.3$ and $\log \nh=22.5$, a soft component equivalent to
%13 per cent of the primary power law would have this effect.
%end bit.

In the \xmm\ performance-verification observations of the Lockman Hole,
\citet{mainieri02} find only one source that requires a soft component
superimposed on a heavily absorbed primary power law. This object \citep[$\#$50
in Mainieri et~al. 2002, and $\#$901 in ][]{lehmann01} has similar 
properties to the objects in our sample that
show an additional soft component: narrow optical emission lines, $z=0.204$ and
$L_{2-10}\sim 10^{43}$~erg~s$^{-1}$. In the 13$^{H}$ Deep Field, 
\citet{page06b}
also find one absorbed source ($\#100$), 
with convincing evidence for an additional soft component. This source has
 narrow emission lines, $z=0.27$ and $L_{2-10} =4\times 10^{42}$~erg~s$^{-1}$.
 \citet{caccianiga04} find 3 heavily
absorbed objects which require additional soft components among their sample of
serendipitous 4.5-7.5 keV selected \xmm\ sources. Again, all have narrow
optical emission lines, $z<0.3$, and $L_{2-10} < 3\times 10^{43}$~erg~s$^{-1}$.
However, \citet{mateos05b} find 3 heavily absorbed narrow-line AGN in the very
deep \xmm\ observations of the Lockman Hole, which have $0.7 < z < 0.8$ and
$3\times 10^{43}$~erg~s$^{-1} < L_{2-10} < 10^{44}$~erg~s$^{-1}$, and which
require additional soft components. The soft component 0.5-2 keV luminosities
are $< 2\times 10^{42}$~erg~cm$^{-2}$~s$^{-1}$ in all 3 cases, similar to the
soft component luminosities found in our sample.

Our results are in stark contrast to the soft X-ray properties reported for
absorbed sources in the \sax\ HELLAS survey: \citet{vignali01} find
that additional soft components, equivalent to between 25\% and 65\% of the
primary unabsorbed power law emission, are required in 6/12 
heavily absorbed ($\nh > 5\times 10^{22}$~cm$^{-2}$) broad-line sources. 
However, as we have already discussed, the $HR2$ selection criterion will
exclude objects with very strong soft components from our sample, and a
typical source at $z=0.3$ with $\nh = 3\times 10^{22}$~cm$^{-2}$, will only be
selected in our sample for a soft component equivalent to $<16$ per cent 
of the primary emission. Nonetheless, it is hard to explain such powerful soft
components, equivalent to greater than 25 percent of the primary emission, in 
terms of scattered emission (see Section \ref{sec:implications}).
One possibility is that these HELLAS broad line AGN
actually posess ionised absorbers, which in the 
hardness ratio analysis performed
by \citet{vignali01}, could not be distinguished from cold
absorption and a strong soft component, as would be the case for 
XMM~J$080625.35$$+$$244326.0$ in our sample. 
For NELGs, \citet{vignali01} report that 2 absorbed 
HELLAS objects with narrow 
emission lines showed evidence for additional soft X-ray components, equivalent
to 5\% and 35\% of the primary power law emission. While the HELLAS source 
with a soft component equivalent to 5\% of the primary power law is comparable
with our findings, 35\% is
larger than any soft component that would be expected in our 
sample by the $HR2$ selection cretierion.

\begin{figure}
\begin{center}
\leavevmode
\psfig{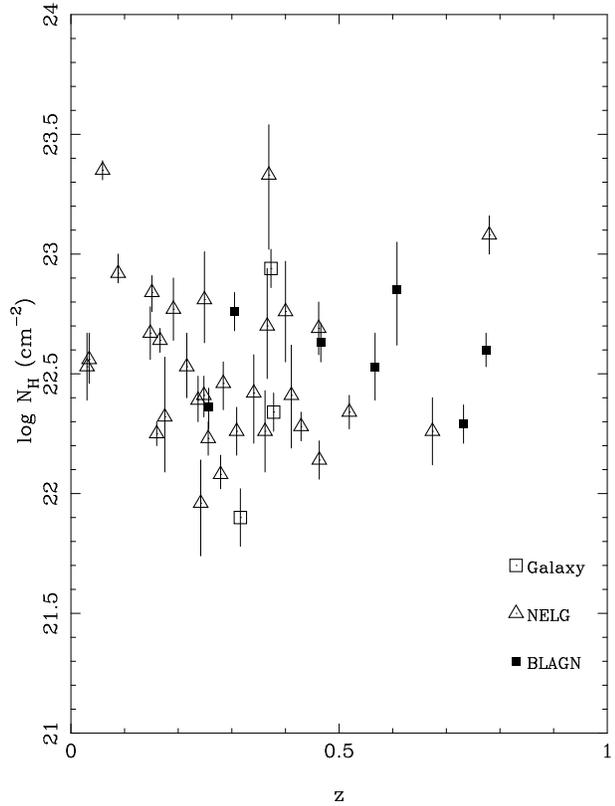}
\caption{Column densities of the sample as a function of redshift.}
\label{fig:nhbestred}
\end{center}
\end{figure}

\subsection{X-ray bright, optically normal galaxies}
\label{sec:discussion_XBONGs}

Three sources in our sample 
(XMM~J$095231.84-015016.1$, XMM~J$133913.92-314421.9$ and 
XMM~J$134656.75+580316.5$)
 have 2-10~keV 
luminosities of $10^{43} - 10^{44}$~ergs~s$^{-1}$, but 
show no emission lines in their optical spectra; they are therefore
``X-ray bright, optically-normal galaxies'' 
\citep[XBONGs; see ][]{comastri02,griffiths95,elvis81}.
As discussed by \citet{severgnini03} and \citet{moran02}, the majority of such 
objects are thought to be AGN in which the emission lines are
undetectable against the host galaxy starlight except in high signal to noise
observations with subarcsecond spatial resolution. As discussed in
\citet{page03a} and \citet{severgnini03}, both absorbed and unabsorbed AGN 
could contribute to the XBONG population. 
Our three sources are 
consistent with the absorbed AGN scenario, as
all three show significant X-ray 
absorption.
% and their optical spectra have lower
%signal/noise ratios than the spectra examined by \citet{severgnini03}.
%xray absorbed
%emission lines swamped by host galaxy light
%our spectra poorer statistical quality than severgnini et al.

\subsection{Limits on the gas-to-dust ratios in BLAGN}
\label{sec:discussion_BLAGN}

Fig. \ref{fig:nhbestred} shows the distribution of absorber $\nh$ 
from the best fit models as a
function of redshift. $\nh$ ranges from 
$7\times 10^{21}$~cm$^{-2}$ to $3\times 10^{23}$~cm$^{-2}$, with a mean of 
$3\times 10^{22}$~cm$^{-2}$. The distributions of $\nh$ for the BLAGN, NELGs
and galaxies are indistinguishable with a KS test.
There is no correlation between log~$\nh$ and redshift 
(linear correlation coefficient of $-0.05$), which is interesting because the
hardness ratio criterion (Equation \ref{eq:selection}) should select objects
with higher column density at higher redshift and higher luminosities. 
The paucity of luminous ($L_{2-10} > 10^{44}$~erg~cm$^{-2}$~s$^{-1}$) sources 
with $\nh > 10^{23}$~cm$^{-2}$ in the sample implies either that such 
sources are rare, or that the optical limits have excluded these objects from 
the sample.

It is rewarding to compare our sample with the sample of
\ros\ sources presented in \citet{page01}, which is also 
a hard-spectrum, X-ray colour selected sample, has a 0.5-2 keV flux limit 
which
is similar to our 2-4.5 keV flux limit, and has a 
similar optical magnitude limit to the sample presented here. Our 
\xmm\ sample is selected at higher energies, so that at any given redshift it
selects objects with higher column densities than the \ros\ sample. Whereas our
\xmm\ sample contains no objects at $z>1$, almost a quarter of the \ros\ 
sample (14/62 objects) have $z>1$, all of which are BLAGN with
$21.4 < \log \nh < 22.6$; most of them show little attenuation by dust in 
their ultraviolet spectra and have $R<21$ \citep[e.g.][]{page00}. At $z>1$ 
the flux limits and X-ray 
colour selection of our survey mean that it is most sensitive to AGN
with $23 < \log \nh < 23.5$.
Therefore the lack 
of BLAGN with 
$\nh > 10^{23}$~cm$^{-2}$, and the absence of any objects at $z>1$
in our $R<21$ \xmm\ sample implies that BLAGN with 
$\nh > 10^{23}$~cm$^{-2}$, {\em but without significant attenuation in the 
UV}, 
are rare.  Assuming that this is because of extinction in the optical/UV, 
we can estimate 
the maximum effective gas-to-dust ratio\footnote{The Galactic gas-to-dust 
ratio which would produce equivalent UV extinction for the X-ray derived cold
column density.} which is found in our X-ray absorbed AGN sample
as follows. To have been 
extinguished beyond the optical limits of the sample ($R=21$), 
BLAGN 
%with $\nh = 10^{23}$~cm$^{-2}$
 at $z\sim 1$ 
would have to be attenuated by more than 1 magnitude at 3500\,\AA\ (restframe)
compared to the \ros\ sample of \citet{page01} in which all but two BLAGN have
$R<20$. This would be equivalent to an extinction of $E(B-V)=0.2$ for Galactic
dust\citep{rieke85}, which would correspond to 
$\nh\sim 10^{21}$~cm$^{-2}$ for a Galactic gas-to-dust ratio \citep{bohlin78}. 
Thus we should observe BLAGN with $z>1$ and $\nh > 10^{23}$~cm$^{-2}$  
in our sample if they have effective gas-to-dust 
ratios $\ge 100 \times$~Galactic.
Since we do not observe such sources, we infer that BLAGN with 
effective gas-to-dust ratios $\ge 100 \times$~Galactic 
are absent from our sample. 

The space density of $z>1$ X-ray absorbed BLAGN with $21.4< \log \nh < 22.6$ is
$\sim$10\% the space density of unobscured BLAGN 
\citep{akiyama00,mainieri02,page04,perola04,silverman05,page06b}. 
The lack of
BLAGN with $23 < \log \nh < 23.5$ in our \xmm\ sample (at higher energies than the
\ros\ sample, but with a similar limiting flux) implies that such objects have
an even smaller space density, equivalent to no more than a few percent of the
unobscured population. It is interesting to compare this to the space density
of broad absorption line (BAL) QSOs, which make up $\sim 15\%$ of the optical
QSO population in the redshift interval ($1.7<z<3.5$) where the BALs can 
be detected
\citep{reichard03}, and approximately half of which have $\nh >
10^{23}$~cm$^{-2}$ \citep{gallagher02}.  BALQSOs are the only subset of the
optically-selected BLAGN population which appear to have effective gas-to-dust
ratios $\sim 100 \times$Galactic \citep{maiolino01}, showing both strong UV
emission and large X-ray column densities.  The limits from our survey imply
that BALQSOs will also form the majority of BLAGN with extreme gas-to-dust 
ratios
amongst the X-ray selected population.
%emission line diagnostics

%luminosities of sources, Seyfert galaxies to QSOs
%soft excess components, vs ionised absorber, cf local type 1 and type 2s.
%luminosities of soft excesses, where they will affect surveys in the future.
%host galaxy luminosities - what types of galaxies, any biases?

\subsection{Implications for the AGN population and geometric unification}
\label{sec:implications}

The phenomonology of X-ray absorbed AGN is an important test of the AGN
unification scheme, which postulates that AGN are
surrounded by a dense obscuring torus of dust and gas \citep[][]{antonucci93},
or optically thick material associated with the outer accretion disc 
\citep[e.g. ][]{elitzur06},
and therefore the X-ray
and optical absorption properties of AGN are primarily determined by their
orientation with respect to the observer. Overall, the results found here and
elsewhere are in broad agreement with this unification scheme, because the
majority of the sources that show significant X-ray absorption
(i.e. $N_{H}>10^{22}$~cm$^{-2}$) are NELGs, and so have optical 
properties consistent
with an edge-on perspective, namely narrow emission lines and absent or
highly-attenuated nuclear UV/optical continua. From the X-ray spectra of 31
such NELGs presented earlier, there are 4 secure detections of an additional
soft component which is equivalent to between 2 and 7 per cent of the primary,
absorbed, emission. In addition, there are 3 tentative detections, and for most
of the sources the upper limits do not rule out a soft component equivalent to
several per cent of the primary emission.  The secure detection rate alone
implies that such components are present in $> 4\%$ of the population
%poisson lower limit 95% on 4 is 1.366 in Gehrels et al.
\citep[at 95\% confidence, based on the estimators of ][]{gehrels86}. Thus a
non-negligible fraction of these heavily absorbed, narrow-line objects show an
additional component of soft X-ray emission that apparently escapes the
obscuring torus.

In the context of the unified
scheme, some soft X-ray emission is expected to arise from obscured
AGN in the form of X-rays that are reprocessed or scattered in the
ionisation cones of the AGN. In such a geometry, the amount of soft
X-ray radiation that can be directed into our line of sight depends on
the opening angle of the torus, and the fraction of the soft X-ray
emission that is absorbed or scattered in the ionisation
cones. Assuming that the torus covers 80\% of the sky, as inferred in optical
and X-ray studies of local AGN \citep{maiolino95,risaliti99}, and that the
radiation that is intercepted within the ionisation cones is scattered
or re-emitted isotropically (so that half is directed back towards the
torus), then the luminosity of the scattered component will be 10
per cent of the luminosity that was absorbed or scattered within the
ionisation cone. Thus to produce a soft component equivalent to 5
percent of the primary power law requires that $\sim$ half of the soft
X-rays are absorbed or scattered within the ionisation cone. For such
a large fraction of the soft X-ray radiation to be intercepted within
the ionisation cone implies that the cone must contain an ionised
absorber with a large column density. 

Thus within the context of the unified scheme, the fraction of
absorbed, narrow line objects that show an additional soft component,
equivalent to several per cent of the primary emission, should be matched by
(at least) the same fraction (i.e. $> 4$ per cent) of broad line objects 
displaying significant
absorption ($\ge 50$ per cent) of their soft X-rays from an ionised
absorber. Note that this is a lower limit to the fraction of heavily absorbed
broad line objects, since ionised absorption close to the base of the
ionisation cone will not redirect X-rays over the edge of the torus.

While many Seyfert 1s in the local Universe have ionised absorbers
\citep{reynolds97,george98,blustin05}, only a few nearby objects such as
NGC\,3783 \citep{blustin02,netzer03,krongold03,behar03} and H\,0557-385
\citep{ashton06} are known to absorb more than half of the incident soft X-ray
emission. Nonetheless, the presence of these 2 objects within the
19 BLAGN in the \citet{piccinotti82} sample is consistent with our expectation
of $> 4$ per cent. For more distant objects, the fraction of BLAGN with heavy
soft X-ray absorption is $\sim 10$ per cent 
\citep{akiyama00,mainieri02,page04,perola04,silverman05,page06b}. This is
compatible with our findings for the NELGs provided that $> 40$ per cent of
these X-ray absorbed BLAGN are viewed pole-on and harbour high-opacity 
ionised 
absorbers, so that $> 4$ per cent of {\em all} BLAGN have this property. 
Evidence that this could be the case is building up: within the 
sample presented here an ionised absorber provides the best fit to
the spectrum of XMM~J$080625.35+244326.0$, and examples of X-ray absorbed 
BLAGN which require ionised absorbers have also been found by 
\citet{piconcelli02}, \citet{page06a}, and \citet{page06b}. Objects 
classified as ``soft X-ray weak quasars'' \citep{brandt00} also appear to be
BLAGN with substantial soft X-ray absorption, and the absorbers in a 
significant fraction of these are also found to be 
ionised \citep{schartel05}. These results indicate that within the context of
the unified scheme, the current limits on
the scattered components in narrow line objects are compatible with the numbers
of X-ray absorbed broad line objects. 

Finally, it is worth asking how the three galaxies without discernable optical 
emission lines fit within this picture. In our sample, they are 
outnumbered $\sim$10:1 by the NELGs, and so are a minority of the absorbed 
AGN population. In trying to fit these objects with the unified scheme, we 
can envisage four
possible hypotheses for the apparent weakness of their optical emission lines:
(i) the AGN are so weak compared to the host galaxies that the emission 
lines are
undetectable against the bright galaxy continua, (ii) the tori in
these objects have covering factors approaching unity, (iii) their narrow
line regions are themselves obscured, or (iv) the gas in their narrow line 
regions is so tenuous as to make the optical emission lines anomalously
weak. Hypothesis (i) can be ruled out immediately: their optical to X-ray
flux ratios lie within 0.5 dex of the $f_{X}/f_{R}=1$ line shown in Fig. 
\ref{fig:X_R}, and are indistinguishable from those of the NELGs. In all the 
remaining possibilities, the fraction of soft X-rays that can be reprocessed
and redirected into our line of sight should be negligible, and so significant 
soft X-ray components in their X-ray spectra would be {\em inconsistent} with 
the unified scheme. Soft components are not detected in any of the 3,
as expected from the unified scheme.   

\section{Conclusions}
\label{sec:conclusions}

We have defined the XMM-SSC survey of hard-spectrum \xmm\ sources, and
presented optical and X-ray spectra for an optically-bright ($r<21$) sample of
42 sources.  We find that our hardness ratio selection criterion provides a
very clean sample of X-ray absorbed AGN, in agreement with the findings of
\citet{caccianiga04} and \citet{dellaceca04}.  Every source in the optically
bright sample is an AGN with $21.8 < \log \nh < 23.4$, with the possible
exception of one source. This source has an optical counterpart which is a
Galactic M star, but we argue that a fainter optical counterpart, closer to the
X-ray source position, is more likely to be the correct optical counterpart.
Most (36/42) of the sources have X-ray luminosities characteristic of Seyferts
($10^{42}$~erg~s$^{-1} < L_{2-10} < 10^{44}$~erg~s$^{-1}$). The majority of the
sources (31/42) show only narrow emission lines (FWHM $<$ 1000 km\,s$^{-1}$) in
their optical spectra, although a small number (7) of the sources also show
broad emission lines, and 3 sources have no detectable emission lines in their
optical spectra.  None of the sources have $z>1$. The lack of broad line AGN at
higher redshifts, luminosities and column densities implies that the effective
gas-to-dust ratio in AGN absorbers rarely exceeds $100 \times$ Galactic, and 
it is
unlikely that more than a few percent of broad line AGN have X-ray column
densities of cold gas which are $>10^{23}$~cm$^{-2}$. In 7 of the 31 NELGs an
absorbed power law is unable to fit the X-ray spectrum satisfactorily; an
additional soft component provides acceptable fits for these objects. In our
sample, these soft components are only observed in narrow emission line objects
which are relatively nearby ($z<0.3$) and have low-moderate luminosity
($L_{2-10}<3\times 10^{43}$~erg~s$^{-1}$).  These soft components are a natural
consequence of the AGN unified scheme if they are soft X-rays that have been
absorbed and re-emitted within the narrow line region. This would require that
$> 4$ per cent of broad line AGN have ionised absorbers with $> 50$ per cent
opacity to soft X-rays, indicating that a substantial fraction of X-ray
absorbed broad-line AGN have ionised absorbers.

\section{Acknowledgments}

Based on observations obtained with XMM-Newton, an ESA science mission with
instruments and contributions directly funded by ESA Member States and NASA.
This research was also based on observations made at the William Herschel
Telescope which is operated on the island of La Palma by the Isaac Newton Group
in the Spanish Observatorio del Roque de los Muchachos of the Instituto de
Astrofisica de Canarias, and on observations collected at the European Southern
Observatory, Chile, ESO No. 71.A-0444. This research has made use of the
NASA/IPAC Extragalactic Database (NED) which is operated by the Jet Propulsion
Laboratory, California Institute of Technology, under contract with the
National Aeronautics and Space Administration. This research has made use of
data obtained from the SuperCOSMOS Science Archive, prepared and hosted by the
Wide Field Astronomy Unit, Institute for Astronomy, University of Edinburgh,
which is funded by the UK Particle Physics and Astronomy Research Council.  FJC
acknowledges financial support from the Spanish Ministerio de Educacion
Ciencia, under projectw ESP2003-00812 and ESP2006-13608-C02-01. 
RDC and PS acknowledge partial financial
support by hte MIUR (Cofin-03-02-23) and INAF.  We thank Natalie Webb for
useful comments.

\end{document}